\newif\ifsubmode
\submodetrue

\newif\ifincludetable
\includetablefalse

\ifsubmode
\documentclass[10pt,preprint]{aastex}
\else
  \documentclass{emulateapj}   \usepackage{apjfonts}
\fi
\newcommand{\hst}{\textit{HST}}
\newcommand{\iras}{\textit{IRAS}}

\newcommand{\spitzer}{\textit{Spitzer}}

\newcommand{\chandra}{\textit{Chandra}}

\newcommand{\ks}{\hbox{$K_s$}}
\newcommand{\AAA}{\hbox{\AA}}

\newcommand{\lsim}{\lesssim}
\newcommand{\gsim}{\gtrsim}

\newcommand{\etal}{et al.}
\newcommand{\eg}{e.g.}

\newcommand{\msol}{\hbox{$M_\odot$}}

\newcommand{\lsol}{\hbox{$L_\odot$}}

\newcommand{\uJy}{\hbox{$\mu$Jy}}
\newcommand{\ujy}{\hbox{$\mu$Jy}}
\newcommand{\lir}{\hbox{$L_{\mathrm{IR}}$}}
\newcommand{\mone}{\hbox{$[3.6\mu\mathrm{m}]$}}
\newcommand{\mtwo}{\hbox{$[4.5\mu\mathrm{m}]$}}
\newcommand{\mthree}{\hbox{$[5.8\mu\mathrm{m}]$}}
\newcommand{\mfour}{\hbox{$[8.0\mu\mathrm{m}]$}}

\newcommand{\zph}{\hbox{$z_\mathrm{ph}$}}

\newcommand{\lbol}{\hbox{$L_\mathrm{bol}$}}

\shorttitle{MID--TO--FAR-IR FLUX DENSITIES OF DISTANT GALAXIES}
\shortauthors{PAPOVICH ET AL.}

\begin{document}

\title{{\it SPITZER} MID--TO--FAR-INFRARED FLUX DENSITIES OF DISTANT GALAXIES\altaffilmark{1}}
\slugcomment{\it Accepted for Publication in the Astrophysical Journal}


\author{\sc Casey~Papovich\altaffilmark{2,3},
Gregory~Rudnick\altaffilmark{4}, Emeric~Le~Floc'h\altaffilmark{2,3,5},
Pieter~G.~van Dokkum\altaffilmark{6}, George~H.~Rieke\altaffilmark{2},
Edward~N.~Taylor\altaffilmark{7}, Lee Armus\altaffilmark{8}, Eric
Gawiser\altaffilmark{6}, Jiasheng~Huang\altaffilmark{9},
Delphine~Marcillac\altaffilmark{2}, and Marijn~Franx\altaffilmark{7} }

\altaffiltext{1}{This work is based in part on observations
made with the \textit{Spitzer Space Telescope}, which is operated by
the Jet Propulsion laboratory, California Institute of Technology,
under NASA contract 1407}
\altaffiltext{2}{Steward Observatory, University of Arizona, 933 N.~Cherry Ave., Tucson, AZ 85721; papovich@as.arizona.edu}
\altaffiltext{3}{Spitzer Fellow}
\altaffiltext{4}{Goldberg Fellow, National Optical Astronomy
Observatories, 950 N.~Cherry Ave., Tucson, AZ 85721}
\altaffiltext{5}{Current Address: Institute for Astronomy, University
of Hawaii, 2680 Woodlawn Dr., Honolulu, HI 96822}
\altaffiltext{6}{Department of Astronomy and Yale Center for Astronomy
and Astrophysics, Yale University, New Haven, CT 06520}
\altaffiltext{7}{Leiden Observatory, Leiden University, PO Box 9513,
2300 RA Leiden, The Netherlands}
\altaffiltext{8}{Spitzer Space Center, MS~220-6, California Institute of
Technology, Pasadena, CA, 91125}
\altaffiltext{9}{Harvard-Smithsonian Center for Astrophysics, 60 Garden Street, Cambridge, MA 02138}



\begin{abstract}
We study the infrared (IR) properties of high--redshift galaxies using
deep \spitzer\ 24, 70, and 160~\micron\ data. Our primary interest is
to improve the constraints on the total IR luminosities, \lir,  of
these galaxies. We combine the \spitzer\ data in the southern Extended
Chandra Deep Field with a \ks-band--selected galaxy sample and
photometric redshifts from the Multiwavelength Survey by
Yale-Chile. We used a stacking analysis to measure the average 70 and
160~\micron\ flux densities of $1.5 < z < 2.5$ galaxies as a function
of 24~\micron\ flux density, X-ray activity, and rest--frame near-IR
color. Galaxies with $1.5 < z < 2.5$ and $S_{24}=53-250$ \ujy\ have
\lir\ derived from their average 24-160~\micron\ flux densities within
factors of 2--3 of those inferred from the 24~\micron\ flux densities
only.   However, \lir\ derived from the average 24--160~\micron\ flux
densities for galaxies with $S_{24} > 250$~\ujy\ and $1.5 < z < 2.5$
are lower than those inferred using only the 24~\micron\ flux density
by factors of 2--10. Galaxies with $S_{24} > 250$~\ujy\ have
$S_{70}/S_{24}$ flux ratios comparable to sources with X-ray
detections or red rest--frame IR colors, suggesting that warm dust
possibly heated by AGN may contribute to the high 24~\micron\
emission. Based on the average 24--160~\micron\ flux densities, nearly
all 24~\micron--selected galaxies at $1.5 < \zph < 2.5$ have $\lir <
6\times 10^{12}$~\lsol, which if attributed to star formation
corresponds to $\Psi < 1000$~\msol\ yr$^{-1}$.  This suggests that
high redshift galaxies may have similar star formation efficiencies
and feedback processes as local analogs.   Objects with $\lir > 6\times
10^{12}$~\lsol\ are quite rare, with a surface density $\sim 30\pm
10$~deg$^{-2}$, corresponding to $\sim 2\pm 1\times10^{-6}$~Mpc$^{-3}$
over $1.5 < z < 2.5$.
\end{abstract}
 
 

\section{Introduction}\label{section:intro}

Observations with \textit{ISO}, SCUBA, and more recently the \spitzer\
Space Telescope show that high redshift galaxies contain large amounts
of dust, which emit strongly at infrared (IR) wavelengths.   Although
locally, IR--luminous galaxies are rather rare
\citep[\eg,][]{soi87,soi91}, number counts from \textit{ISO} and
\spitzer\ demonstrate that this population evolves very rapidly
\citep{elb99,pap04}, dominating the star--formation rate (SFR) density
and cosmic IR background by $z\sim 1$
\citep[\eg,][]{fra01,elb02,lef05,per05,dol06}.     Recent observations
from the Multiband Imaging Photometer for \spitzer\ (MIPS; Rieke et
al.\ 2004)  are sensitive to the IR emission from galaxies at yet
higher redshifts.  Several studies using
\spitzer\ 24~\micron\ observations demonstrated that at $1.5 \lsim z
\lsim 3$ the density of ultra--luminous IR galaxies with
$L(8-1000\micron) > 10^{12}$~\lsol\ (ULIRGs) was $\approx 1000\times$
higher than at present \citep{dad05,pap06}, and that $\sim 50$\% of
massive galaxies (stellar masses, $M \gsim 10^{11}$~\msol) at these
redshifts emit strongly at 24~\micron\
\citep{dad05,cap06,pap06,red06,web06}.   Thus, more than half of the
massive galaxies at $z$$\sim$2 are either actively forming stars,
fueling supermassive black holes, or both.

The high incidence of IR--active, massive galaxies at $1.5 \lsim z
\lsim 3$ coincides with rapid evolution in the stellar mass density.  
Most ($\gsim$50\%) of the stellar mass in galaxies today formed during
the short time between $z$$\sim$3 and 1
\citep{dic03,rud03,rud06,fon03,fon04,gla04}.   Much of this stellar
mass density resides in massive galaxies, which appear at epochs prior
to $z$$\sim$1--2 \citep[see][for reviews]{mcc04,ren06}.   This is
consistent with recent theoretical work by, for
example, \citet{delucia06} who argue that most of the star--formation
in massive galaxies occurs at early lookback times.

The \spitzer\ IR observations of high redshift galaxies may be
revealing vigorous star formation episodes.   However, to date, most
studies rely on converting the 24~\micron\ flux densities from
\spitzer\ to total IR luminosities, $\lir \equiv L(8-1000\micron)$,
and then to the instantaneous SFR \citep[\eg,][]{ken98}.  At $z\sim 2$
the 24~\micron\ band probes rest--frame 8~\micron.  While this
wavelength broadly correlates with \lir, there may be large variations
with bolometric luminosity and galaxy type
\citep[\eg,][]{cha01,rou01,elb02,cal05,cal07,alo06b,bra06}, and significant scatter
may be expected because of the range of shapes of the IR spectral
energy distributions (SEDs) \citep[\eg,][]{dal01,dal05,pap02,arm06}.  

Broadly, there are few constraints on the relation between the
observed 24~\micron\ (rest--frame 8~\micron) and the total bolometric
emission for $z\sim 2$ galaxies. 
\citet{dad05} found that the average (stacked) X-ray, UV, mid--IR,
sub--mm, and radio emission of $BzK$--selected star-forming galaxies
at $1.4 < z < 2.5$ gave consistent estimates for the instantaneous
SFR.  
\citet{ega04} and \citet{pop06} showed that the 24~\micron\ emission from $z\sim 2-3$
sub-mm--selected objects provided an accurate measure of the total
bolometric emission for galaxies relative to that inferred from the
combined 24~\micron, 850~\micron, and 1.4~GHz radio data.   Although
\textit{on average} 24~\micron\ observations of $z\sim 2$ provide an
accurate measure of the total IR luminosity (at the factor 2 level),
one must exercise care when using 24~\micron\ data as a total IR
indicator at high redshifts as the conversion between the mid--IR and
total IR have not yet converged \cite[\eg,][]{pap02,arm06,dad07a}, and
individual estimates may have large errors.

Observations at wavelengths longer than 24~\micron\ are required to
improve our constrains on the shape of the IR spectral energy
distribution (SED) in distant galaxies.   At redshifts $z$=2
the \spitzer\ 70~\micron\  and 160~\micron\ bands probe rest--frame 24
and 55~\micron, respectively.  The rest--frame 24~\micron\ emission, in
particular, correlates strongly with the total IR luminosity and SFR
with substantially smaller scatter than the luminosity at rest--frame
8~\micron\ for star-forming regions and starburst galaxies
(Calzetti et al.\ 2005; Alonso--Herrero \etal\ 2006b; but see Calzetti
et al.\ 2007).  As discussed in
\citet{pap02} the $S_{70}/S_{24}$ ratio better correlates with the total far--IR emission at these
redshifts, providing tight constraints on the total far--IR emission
to $<30$\%.   Furthermore, \citet{sie06} showed that fitting their
theoretical models for the IR emission from galaxies to the
rest--frame 8 and 24~\micron\ flux densities provided a tight
constraint on the shape of the IR SED and total IR luminosity.

In this paper we study the IR properties of high--redshift galaxies
using deep \spitzer\ 24, 70, and 160~\micron\ data.    Our primary
interest is to improve our constraints on the total IR luminosities
from high--redshift galaxies detected at 24~\micron\ using the longer
wavelength MIPS data.   However, 70 and 160~\micron\ observations
detect few  galaxies directly at $z\gsim
1$ \citep[\eg,][]{fra06a,fra06b,huy07,dad07a} owing primarly to the larger
angular resolution and poorer sensitivity of
\spitzer\ at 70 and 160~\micron\ relative to 24~\micron.  
Therefore,  we use stacking methods to study the average 70 and
160~\micron\ flux densities of galaxies.   Here, we study a large
sample of galaxies with $1.5 < z < 2.5$, which allows us to measure
the average 70 and 160~\micron\ flux densities of sub--populations of
galaxies, divided  as a function of 24~\micron\ flux density, and for
galaxies with putative AGN inferred from X-ray activity or red
rest--frame near--IR colors.    Thus, we study the IR properties of
distant galaxies during the epoch where a substantial fraction of
massive galaxies are in active IR phases of their evolution. 

Throughout this work we quote optical and near--IR magnitudes on the
AB system where $m_\mathrm{AB} = 23.9 - 2.5 \log( f_\nu/1\;\uJy)$
unless otherwise specified.    We denote magnitudes measured from the
data with \spitzer\ IRAC in the four channels  [3.6], [4.5], [5.8],
[8.0], respectively.   Similarly, we denote the flux denisty, $f_\nu$,
in the MIPS 24, 70, and 160~\micron\ bands as $S_{24}$, $S_{70}$, and
$S_{160}$, respectively.  To derive rest--frame quantities, we use a
cosmology with $\Omega_m = 0.3$, $\Lambda = 0.7$, and $H_0 = 70$~km
s$^{-1}$ Mpc$^{-1}$. 

\section{Data}\label{section:data}

For this study, we used data in the Extended \textit{Chandra} Deep
Field South (ECDF--S), which has field center coordinates
$3^\mathrm{h}32^\mathrm{m}30^\mathrm{s}$,
$-27^\circ48^\prime20^{\prime\prime}$.   The ECDF--S has been targeted
by a large array of ground--based and space--based telescopes, and it
has deep multiwavelength coverage.   For this study, we make use of
datasets covering a large area (775 arcmin$^2$) in the ECDF--S from
the Multiwavelength Survey by Yale--Chile (MUSYC), the
\spitzer\ Space Telescope, and the \chandra\ X-ray Observatory. 

\subsection{MUSYC}

The MUSYC data include $UBVRIzJH\ks$ imaging in a $31^\prime \times
31^\prime$ field in the ECDF--S. \citet{gaw06} discuss the
observations, images,  and data reduction of the
$U$--through--$z$-band images.   E.~Taylor et al.\ (2007, in
preparation) describe the details of the $JH\ks$ observations  and
data reduction, and of the source detection and cataloging processes.
In summary, we performed object detection and photometry on images
convolved to match the image quality of the  image with the poorest
seeing to account for seeing variations.   Objects were detected in
the $\ks$--band data using the SExtractor software \citep{ber96}, and
colors were measured in $2\farcs25$--diameter apertures (SExtractor
MAG\_APER magnitudes) on the seeing--matched images in each band.  We
scaled the aperture magnitudes to total magnitudes using the
difference between the \ks--band aperture (MAG\_APER) and total
(MAG\_AUTO) apertures, $\Delta m = \mathrm{MAG\_APER}(\ks) -
\mathrm{MAG\_AUTO}(\ks)$.    The catalog reaches a $5\sigma$ sensitivity
of  $\ks = 22.2$~mag (E.~Taylor et al.\ 2007, in preparation)

\subsection{\spitzer}

The Spitzer Guaranteed Time Observers (GTOs) observed the ECDF--S with
both the IRAC and MIPS instruments, covering $3-160$~\micron.   IRAC
data covers 3--8~\micron\ in four bands centered at 3.6, 4.5, 5.8, and
8.0~\micron.  We detected objects using SExtractor on the IRAC data
using a weighted--summed \mone+\mtwo\ image.  We then performed
photometry using SExtractor in circular apertures of diameter
4\arcsec, and converted these to total magnitudes by applying aperture
corrections derived from the measured point source curve--of--growth
of 0.30, 0.34, 0.53, and 0.67~mag to the \mone, \mtwo,
\mthree, and \mfour\ photometry,  respectively.  
To estimate the photometric uncertainties we repeatedly added
artificial sources to the IRAC data, and reperformed object detection
and photometry.    From our analysis of these simulations the
$5\sigma$ magnitude limits for point--sources are  23.3, 22.7, 21.3,
and 21.6~mag in \mone, \mtwo, \mthree, and \mfour, respectively.

\begin{deluxetable}{lccccc}
\tablewidth{0pt}
\tablecaption{Flux Completeness and Accuracy of the ECDF--S
\spitzer/MIPS Data\label{table:mips_data}}
\tablehead{ \colhead{Band} & \colhead{ $\langle t_\mathrm{exp}\rangle$} & 
\colhead{$C(50\%)$} & \colhead{$C(80\%)$} &
\colhead{$\langle f_\nu(\mathrm{S/N}=3)\rangle$} & 
\colhead{$\langle f_\nu(\mathrm{S/N}=5) \rangle$} \\
\colhead{(1)} & \colhead{(2)} & \colhead{(3)} & \colhead{(4)} &
\colhead{(5)} & \colhead{(6) }}
\startdata
24~\micron & 2700~s & 51\phd\phn & 71\phd\phn & 53\phd\phn & 110\phd\phn \\
70~\micron & 1100~s & \phn3.9 & \phn6.6 & \phn 4.6 & \phn\phn8.2 \\ 
160~\micron & 300~s & 20\phd\phn & 44\phd\phn & 24\phd\phn & \phn59\phd\phn
\enddata
\tablecomments{Units of flux density are \ujy\ for 24~\micron\ and mJy
for 70 and 160~\micron.  }
\end{deluxetable}

\citet{pap04} and \citet{dol04} describe the data reduction and
point--source photometry methods applied to the \spitzer/MIPS 24, 70,
and 160~\micron\ images, following the procedures described in
\citet{gor05}.    Here, we use rereduced versions of the MIPS images
that combine data from the original observations covering $0.5^\circ
\times 1^\circ$ with a second epoch observation, doubling the exposure
time in an area $\approx$$0.5^\circ \times 0.5^\circ$ centered on the
ECDF--S field.   Table~\ref{table:mips_data} gives the average
exposure time for MIPS in the combined observations.  The
second--epoch observations also reduce data artifacts, particularly in
the 70 and 160~\micron\ images.    We have reanalyzed these data using
the simulations described in \citet{pap04} to estimate the photometric
errors and source completeness in the MIPS images.
Table~\ref{table:mips_data} gives the 50 and 80\% flux--density
completeness limits, $C$(50\%) and $C$(80\%),  and the average flux
densities of sources with S/N=3 and 5 as derived from the simulations.
The sample of 70 and 160~\micron\ sources with 3$<$S/N$<$5 will
include some spurious detections.  Nevertheless, we require that these
70 and 160~\micron\ have 24~\micron\ counterparts, which improves the
reliability of these detections and provides upper limits on
$S_{70}/S_{24}$ and $S_{160}/S_{24}$ flux ratios.

We matched sources in the IRAC and MIPS catalogs to the MUSYC
$\ks$--band catalog down to their approximate $3\sigma$ limits
(53~\uJy, 4.6~mJy, and 24~mJy for 24, 70, and 160~\micron\
respectively).  We used a wavelength--dependent matching radius.
For IRAC, we matched sources with $r \leq 1\arcsec$ to
\ks--band counterparts.  Similarly for MIPS sources, we matched
objects with $r\leq 1$, 4, and 16$\arcsec$ to \ks--band counterparts
for 24, 70, and 160~\micron\ sources, respectively.  The larger
matching radii at 70 and 160~\micron\ allow for centroid shifts owing
to the lower resolution and source confusion \citep[see,
\eg,][]{hog01}.  However, the larger matching radii increase the
likelihood of multiple sources associated with the MIPS source.
Therefore, we require that the 70 and 160~\micron\ sources have
24~\micron\ counterparts, improving the confidence in the
associations. 

\subsection{\chandra}

Two datasets with \chandra\ exist in the ECDF--S.  The observations
include a deep, 1~Msec, central field covering $\approx
400$~arcmin$^2$ \citep{gia02,ale03} and a four--pointing mosaic with
250~ksec depth covering $\approx$1000~arcmin$^2$ over the ECDF--S
\citep{leh05,vir06}.    For this study, we use both the deep \chandra\
catalogs from \citet{ale03} and the shallow--wide-area catalogs from
\citet{leh05}.  The deep
\chandra\ data have aim--point flux limits (S/N$=3$) in the
0.5--2.0~keV and 2--8~keV bands of $\approx 2.5\times10^{-17}$ and
$\approx 1.4\times10^{-16}$~erg~cm$^{-2}$~s$^{-1}$, respectively.  The
shallower data over the ECDF--S have aim--point flux limits (S/N$=3$)
in the 0.5--2.0~keV and 2--8~keV bands of $\approx 1.1\times10^{-16}$
and $\approx 6.7\times 10^{-16}$~erg~cm$^{-2}$~s$^{-1}$, respectively.
Assuming an X-ray spectral slope of $\Gamma=$~2.0, a source detected
with a flux of $\approx 10^{-16}$~erg cm$^{-2}$ s$^{-1}$ would have
both luminosities of $\approx 5.2\times 10^{41}$~erg~s$^{-1}$ and
$\approx 7.6\times 10^{42}$~erg~s$^{-1}$ in either the soft or hard
X-ray band at $z=1$ and $z=3$, respectively (for this particular
choice of $\Gamma$), assuming no Galactic absorption.  The median
positional accuracy for the sources in the \chandra\ catalogs is
$0\farcs6$.   We matched the \chandra\ sources from both the deep and
shallow surveys to sources in the \ks--band catalog within $r\leq
1\arcsec$.

\subsection{Redshift Distribution of \spitzer\ Sources}

We used the MUSYC $UBVRIzJH\ks$ data to derive photometric redshifts
for all objects in the \ks--band catalog, following the method of
\citet{rud01,rud03}.  Because we are predominantly interested in the
IR properties of high--redshift galaxies, we are forced to utilize
photometric--redshift techniques as most massive galaxies at high
redshifts are too faint for optical spectroscopic followup
\citep{vandok06}.   Here, we do not include the \spitzer\ data in the
photometric redshift estimates in order to avoid uncertainties in the
stellar population models that may affect the rest--frame near--IR
emission \citep[see the discussion in, \eg,][]{mar06,vanderwel06},  or
possible dust emission at rest--frame wavelengths $>$3~\micron, which
is not included in the galaxy templates used for the redshift
determination.   A comparison between the photometric redshifts,
$\zph$, and available spectroscopic redshifts gives an accuracy of
$\Delta(z) / (1+z) \simeq 0.06$ for galaxies at $z\leq 1.5$ using a
biweight location estimator (see Rudnick et al.\ 2006).  For dusty
star-forming galaxies at $z > 1.5$ the accuracy is
$\Delta(z)/(1+z)\simeq 0.1$, or $\Delta(z) = 0.3$ at $z=2$.   A more
detailed discussion of the photometric redshifts is given elsewhere
(E.~Taylor et al.\ 2007, in preparation).

\begin{figure}[th]  
\epsscale{1.12}
\plottwo{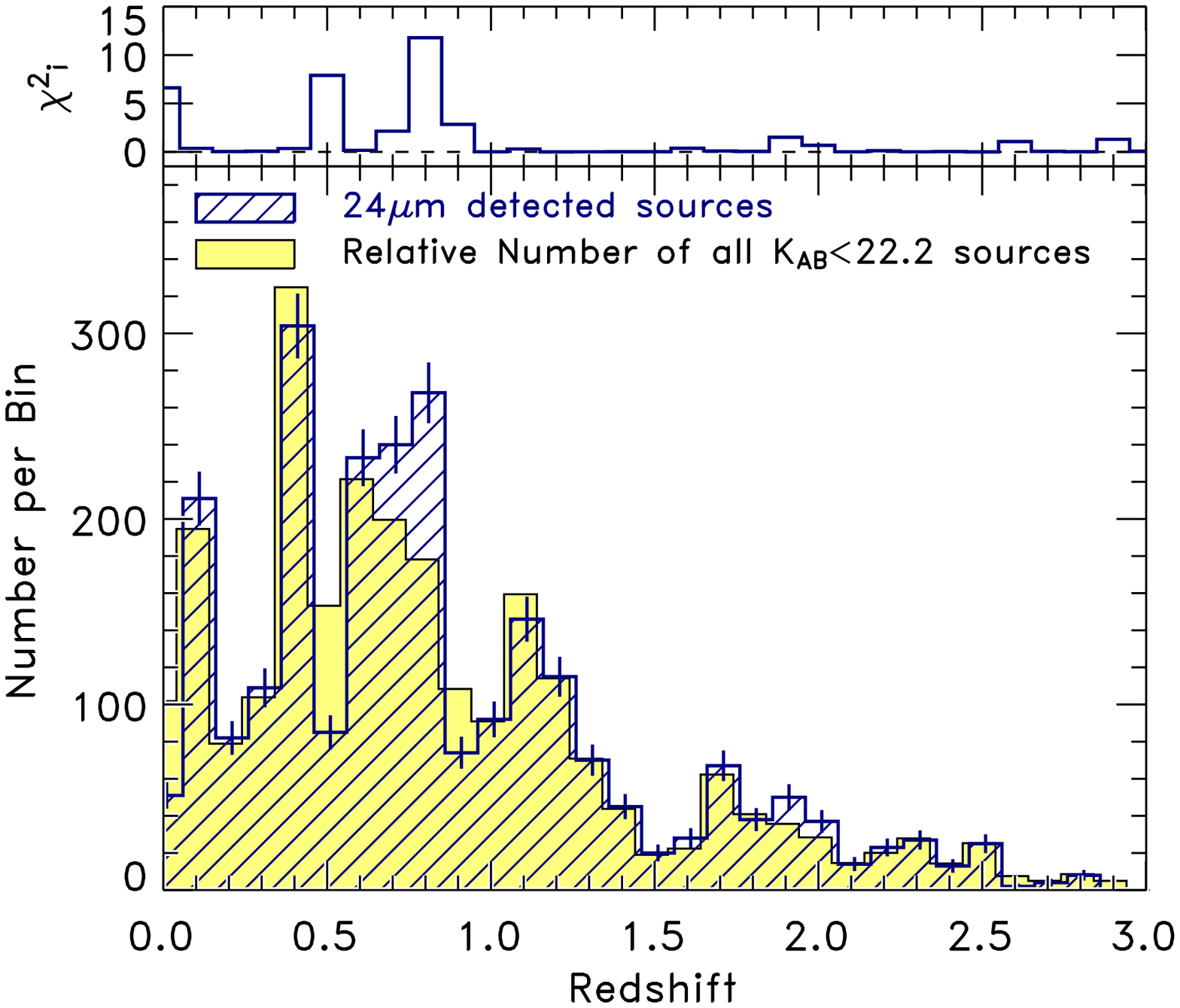}{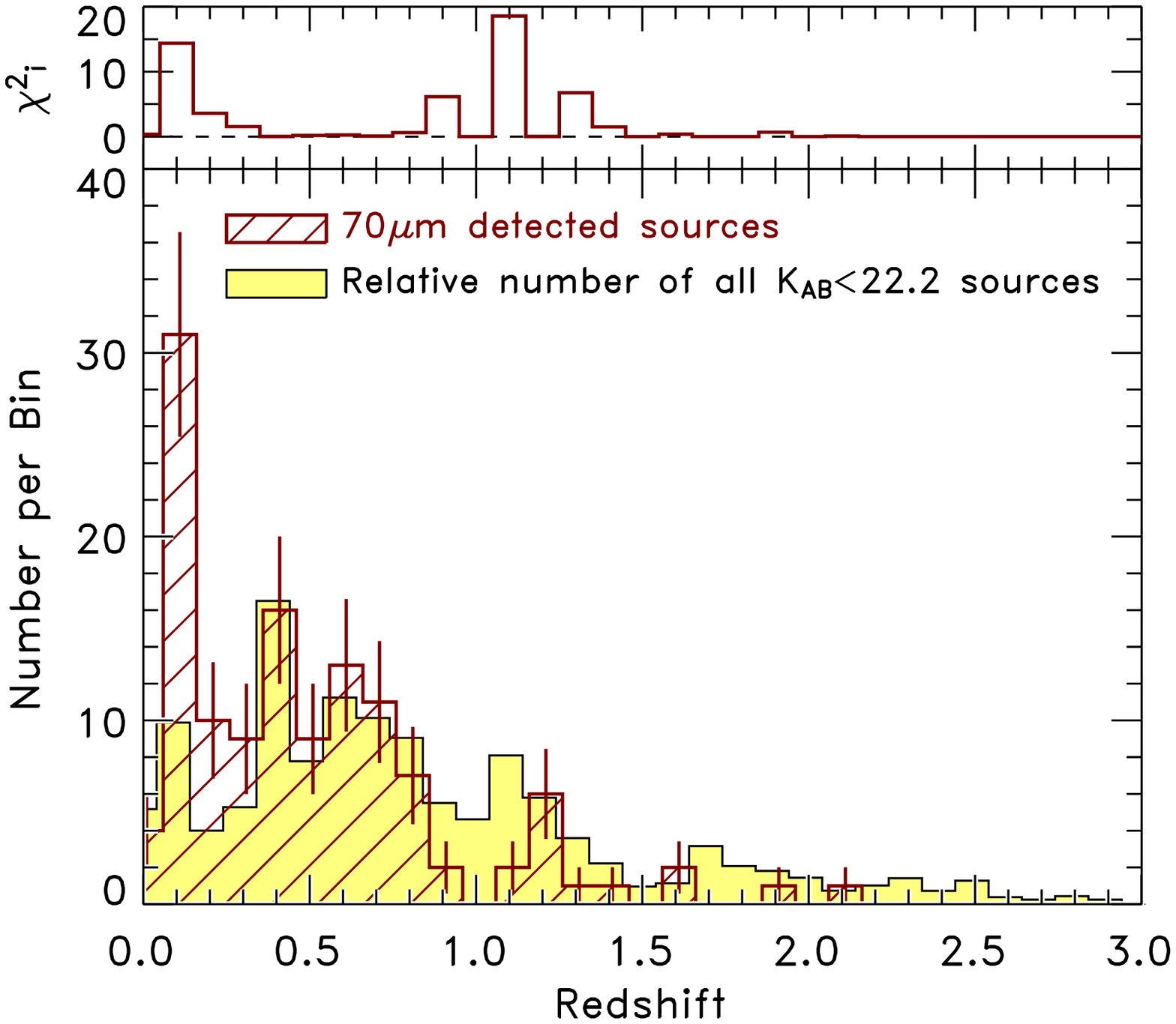}  
\epsscale{1.00}  
\caption{Redshift distribution of MIPS--selected
galaxies in the ECDF--S field.  The left panel shows the redshift
distribution of 24~\micron--selected galaxies with $S_{24} > 53$~\ujy\
and $\ks<22.2$~mag.  The right panel shows the redshift distribution
of 70~\micron--selected galaxies with $S_{70} > 4.6$~mJy and $\ks <
22.2$~mag.   The error bars show the Poissonian uncertainties on the
MIPS distributions only.   In both panels, the yellow, shaded region
shows the relative redshift number distribution of all MUSYC sources
with $\ks < 22.2$~mag, normalized to the total number of MIPS
sources.  The hashed regions shows the redshift number distribution of
MIPS 24 and 70~\micron\ sources (left and right panels, respectively).
In each panel, the upper sub--panel shows $\chi^2_i =
(N^i_\mathrm{exp}-N^i_\mathrm{obs})^2 / N^i_\mathrm{exp}$ for each bin
$i$.  \label{fig:zdist}}
\end{figure}

The MIPS 24, 70, and 160~\micron\ samples matched to the MUSYC
\ks--band catalog have source densities of $13900\pm250$, $740\pm59$,
and $260\pm34$ deg$^{-2}$, respectively, to the $3\sigma$ limiting
flux densities in table~\ref{table:mips_data}.  The median redshifts
of the 24, 70, and 160~\micron\ samples detected at $>$3$\sigma$ are
$z_\mathrm{ph,med} = $0.82, 0.44, and 0.25, respectively.
\citet{fra06a} report similar redshifts for MIPS 70 and 160~\micron\
data in the \spitzer\ first look survey.   Figure~\ref{fig:zdist}
shows the redshift distribution of the 24 and 70~\micron\ sources
matched to the \ks--band selected sample.   The matched
160~\micron--\ks-band list includes only 55 sources, and a histogram
of their photometric redshifts yields little information in addition
to the median quoted above.   The shaded histogram in each panel
shows the number distribution of all \ks--band sources with $\ks
\leq 22.2$~mag, normalized to match the total number of MIPS
sources.   The hashed histograms show the MIPS redshift distributions.
In each figure, the upper panel shows the $\chi^2_i$ per bin $i$,
where $\chi^2_i = (N_\mathrm{exp}^i - N_\mathrm{obs}^i)^2 /
N_\mathrm{exp}^i$, and $N_\mathrm{obs}^i$ is the number of MIPS
sources per redshift bin and $N_\mathrm{exp}^i$ is the expected number
if MIPS sources have the same redshift distribution as the total $\ks
< 22.2$~mag population.

The redshift distribution of the 24~\micron--detected galaxies is
nearly identical to that expected from the \ks--band redshift
distribution.   Although the distributions are similar, a KS test
gives a low likelihood (0.1\%) that the distributions have identical parent
samples.   Qualitatively, there are several interesting deviations.
The strongest relative deficit of 24~\micron\ sources occurs at $z\sim
0.5$.   This may imply that there are relatively fewer IR--active
galaxies at these redshifts in the ECDF--S, or it may result as the IR
emission probed by the 24~\micron\ band shifts from very small grains
to aromatics.  We also observe a relative increase in the number of
24~\micron\ sources at $z\sim 0.7$.   At this redshift the emission
from strong polycyclic aromatic hydrocarbons (PAHs) at
11.3--13.5~\micron\ shift into the 24~\micron\ passband
\citep[e.g.,][]{smi07}, possibly boosting the number of detected
sources at this wavelength. However, we see no evidence for an
increase in relative number of 24~\micron\ sources at $1.5 < z < 2$.
At these redshifts the strong PAH emission feature at 7.7~\micron\
emission shifts into the MIPS 24~\micron\ bandpass.  This contrasts with the
conclusion of \citet{cap06}, who find an relative excess of 24~\micron\ sources at
these redshifts in the GOODS--S field.  Some of this disagreement
may arise from small number statistics as the MUSYC ECDF--S field
encompasses an area $\approx$6 times the size of the GOODS--S field.
However, the GOODS--S data used by Caputi et al.\ extends to a deeper
$\ks$--magnitude limit, and we can not rule out that a relative excess
of 24~\micron\ sources at $1.5 < z < 2.0$ exists for fainter sources.

The redshift distribution of 70~\micron\ sources does not generally
follow the relative distribution of the total \ks--band population.
Formally, a $\chi^2$--test on the distributions shows they differ with
high significance ($>$$4\sigma$).  Because the redshift distribution
of 24~\micron\ sources matches closely the total redshift
distribution, we interpret this as evidence that the  70~\micron\ data
is not sensitive to the IR emission from most high--redshift sources.
That is, at $z\gsim 1$ most IR--active galaxies have 70~\micron\ flux
densities $\lsim 5$~mJy.  This is consistent with a study by
\citet{dye07}, who found that sources with $S_{24} > 70$~\ujy\ and
$z\sim 1$ have flux--weighted average 70~\micron\ flux densities $<
2$~mJy.

\section{Spitzer mid--to--far-IR Colors of Distant Galaxies}

The MIPS flux density ratios (or ``colors'') of galaxies allow us to
study the properties of the dust emission of high--redshift galaxies
using these data.  Because the data detect few 160~\micron\ sources,
we do not discuss the $S_{160}/S_{24}$ or $S_{160}/S_{70}$ ratios for
individual sources here \citep[see, \eg,][]{fra06a}.   We make further
use of the 160~\micron\ data in the analysis of the average flux
densities of high redshift galaxies in \S~\ref{section:stacking}.

\subsection{The \spitzer\ $S_{70}/S_{24}$ Flux Density Ratio as a Function of Redshift}

\begin{figure}[thp] 
\epsscale{0.9} 
\plotone{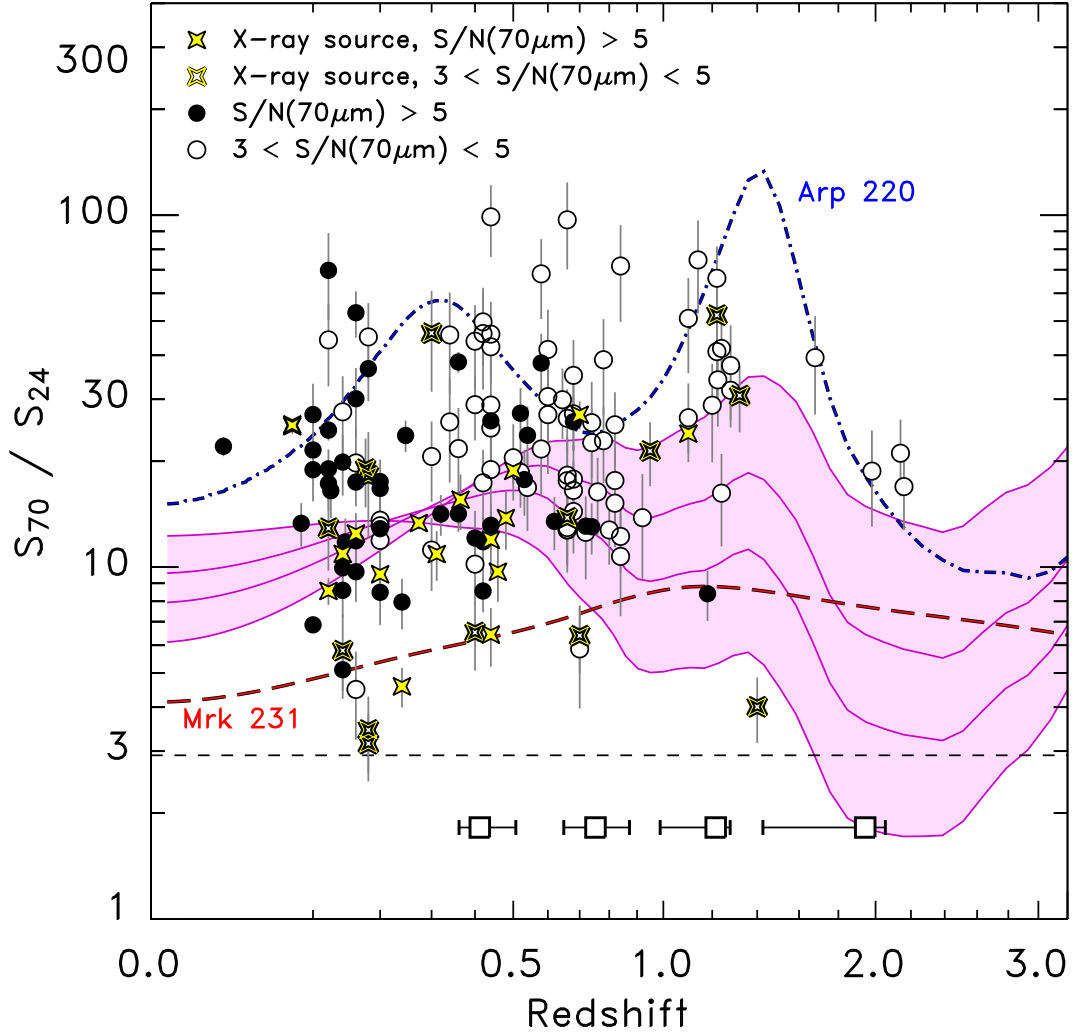}
\epsscale{1.00} 
\caption{MIPS 70~\micron\ to 24~\micron\ flux--density
ratios ($S_{70}/S_{24}$) of galaxies detected at 70~\micron\ in the
ECDF--S as a function of redshift.   Circles denote 70~\micron\
sources without X-ray detections, while stars denote 70~\micron\
sources with X-ray detections.   Sources with  S/N($S_{70}/S_{24}$)
$<$ 3 have been removed for clarity.   The curves show expected
flux--density ratios of galaxy template SEDs.   These include
templates for the local ULIRGs Arp~220 (blue, dot--dashed line) and
Mrk~231 (red, long--dashed line), and empirical templates from Dale \&
Helou (2002; magenta shaded region).   The horizontal short--dashed
line shows the expected flux ratio for a source with constant power in
$\nu f_\nu$, $S_{70}/S_{24}$ $= 2.9$.  The horizontal error bars show
the average uncertainty on the photometric redshifts in various redshift
bins.\label{fig:f70f24vz}}
\end{figure}

Figure~\ref{fig:f70f24vz} shows the $S_{70}/S_{24}$ ratios of the
ECDF--S sources detected at 70~\micron, compared against expectations
from local--galaxy templates.   At $\zph < 1.5$  70~\micron\ sources
with high S/N detections have $S_{70}/S_{24}$ ratios within the bounds
presented by the models, and range from cold--dust dominated ULIRGs
such as Arp~220 to ULIRGs with ``warm'' mid--IR colors such as Mrk~231
\citep{arm06}.   Several objects have $S_{70}/S_{24}$ ratios greater
than that expected from Arp~220.   The most deviant of these outliers
have larger uncertainties on their 24~\micron\ flux density
measurements (3$<$S/N$<$4), reflected in their larger error bars in
the figure.  Several galaxies have $S_{70}/S_{24}$ ratios lower than
that of Mrk~231.  However, these sources all have X-ray detections
(including two that have $S_{70}/S_{24}$ ratios consistent with
constant power in $\nu f_\nu$), suggesting they harbor an AGN that
contributes to the mid--IR emission.

Most of the galaxies (13/15) at $1 < \zph < 1.5$ detected at 70~\micron\ have
high $S_{70}/S_{24}$ ratios, but consistent with the local templates.
These may have strong silicate absorption bands at
$\sim$9--10~\micron\ (like Arp~220; see Armus et al.\ 2007),
suppressing the 24~\micron\ flux density and boosting the
$S_{70}/S_{24}$ ratio.   The remaining 70~\micron\ sources in this
redshift range have $S_{70}/S_{24}$ ratios consistent with
star-forming galaxy templates (Dale \& Helou 2002), not including one
source with a $S_{70}/S_{24}$ ratios consistent with Mrk~231, and one
X-ray source discussed above.

At $1.5 < \zph < 2.5$, four sources are detected directly at
70~\micron.   All four have $S_{70}/S_{24} > 15$, consistent with the
upper envelope of local ULIRG templates.  However, in all cases the
sources have $3 < $ S/N(70\micron) $ < 4$.   Given the relative lack
of 70~\micron\ sources at $z \gsim 1.5$, we reexamined the matching of
the MIPS data to the MUSYC data as well as the galaxy colors ($U-\ks$
and IRAC 3--8~\micron) to verify that the photometric redshifts are
reasonable.   (We remind the reader that the photometric redshifts use
only the MUSYC optical/near--IR data.  For three of the four objects the
70~\micron\ source matches the source in the MUSYC catalog, and the
combined MUSYC and IRAC spectral energy distribution supports the
assigned redshift.   The remaining source lies outside the IRAC GTO
coverage, but within recent \spitzer/IRAC imaging (PI: P.~van Dokkum).
The IRAC data for this source (M.~Damen, 2006, private communication)
supports the photometric redshift, with an SED similar to that of
Arp~220, but with increased extinction in rest-frame optical and UV
wavelengths.  This object is also detected at 160~\micron, and may be
similar to the source studied by \citet{lef06}.    Regardless, the
surface density of $S_{70} > 4.6$~mJy sources at $1.5 < z < 2.5$ is
very low, $\Sigma
\leq 19\pm 9$ deg$^{-2}$.

\subsection{Average Spitzer 70 and 160~\micron\ Flux Densities of $1.5
< z < 2.5$ Galaxies}\label{section:stacking} 

The 70~\micron--detected sources at $1.5 < z < 2.5$ represent only the
``brightest'' galaxies at these redshifts and wavelengths.  To study
the typical 70 and 160~\micron\ flux densities of all 24~\micron\
galaxies at these redshifts, we resorted to stacking techniques to
improve the effective depth of the IR data.  By doing this we lose the
ability to study galaxies on an object--by--object basis, but gain the
ability to study the global trends on the whole.  Stacking  techniques
have already proven valuable to study the IR emission from faint
galaxies \citep{zhe06,zhe07} and to study the contribution of
8~\micron, 24~\micron, and sub--mm selected galaxies to the cosmic IR
background \citep{dol06,dye07,huy07}.

We selected all galaxies from the MUSYC \ks-band
catalog with $S_{24} \geq 53$~\ujy\ and $1.5
\leq \zph \leq 2.5$.    This sample has a mean redshift of $\langle z
\rangle = 1.9$, but the distribution is skewed toward sources with
lower redshift (as is evident in figure~\ref{fig:zdist}).  The
interquartile of the redshift distribution (including the inner
25--75\% of the distribution) spans $1.7 < z_\mathrm{phot} < 2.3$.

We describe the details of our stacking method in the Appendix
section.   Based on our simulations (described in the
Appendix section) we stacked sources in 70 and 160~\micron\ images
first cleaned of sources detected at $>$5$\sigma$.  We also excluded
objects near the edge of the image and a few objects in noisy
(low--weight) regions of the 70~\micron\ image.   The resulting sample
includes 395 galaxies.   We split this sample into 
 bins of 24~\micron\ flux density, $S_{24} > 250$~\ujy, $100 <
S_{24}/\mathrm{\ujy} \leq 250$, and  $53 \leq S_{24} \leq 100$~\ujy. 
We also divide these samples into subsamples with
X-ray detections (45 objects), and objects with red, rest--frame
near--IR colors (76 objects, hereafter ``IR power-law'' sources, see
below).   We hereafter denote those objects without X-ray detections
or red, near--IR colors as ``Ordinary IR sources'' (274 objects).

\begin{figure}[tp]
\plotone{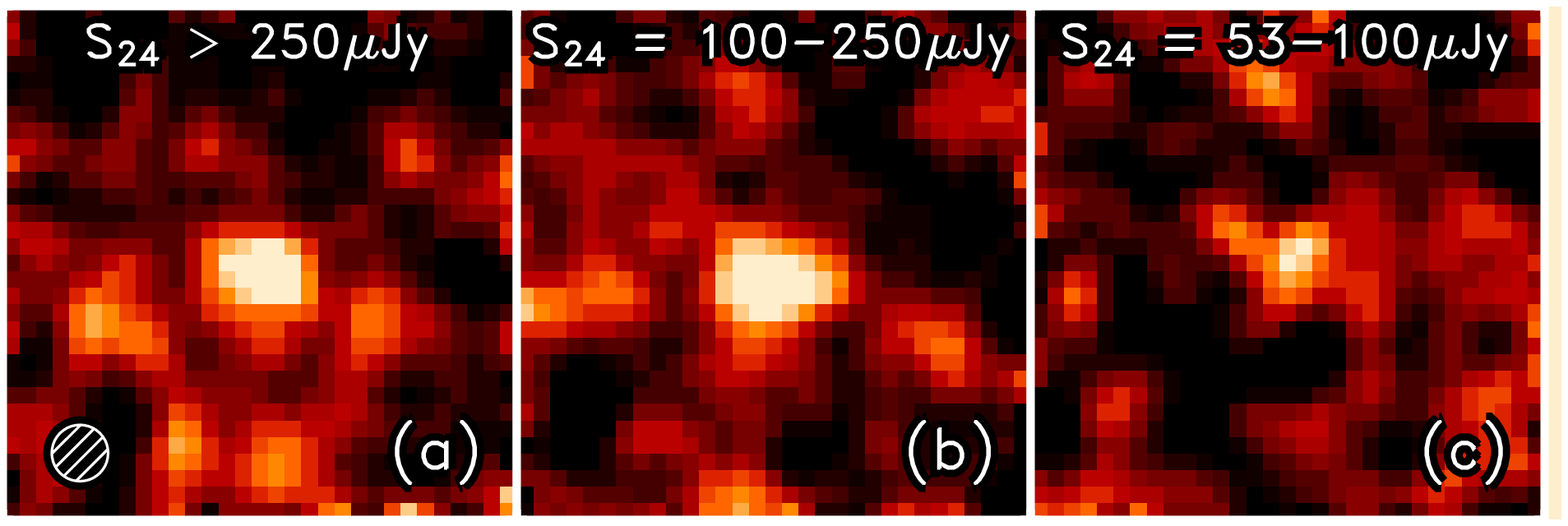}
\caption{ Stacked 70~\micron\ images of ordinary IR sources
at $1.5 < \zph < 2.5$ (i.e., those sources without X-ray detections
and IR--power-law-like colors). The panels show the stacked
70~\micron\ emission in  bins 24~\micron\ flux densities of  (a)
$S_{24} > 250$~\ujy, (b) $100 \leq S_{24}/\ujy < 250$, (c) $53 <
S_{24}/\ujy < 100$.  Each panel shows a region approximately
100\arcsec\ $\times$ 100\arcsec\ (roughly 6$\times$ the 70~\micron\
PSF FWHM), at a scale of $4\farcs925$~pix$^{-1}$. The filled circle,
inset in panel (a), has a diameter equal to the PSF
FWHM.\label{fig:stackedimage}}
\end{figure}
\begin{figure}[bp]
\plotone{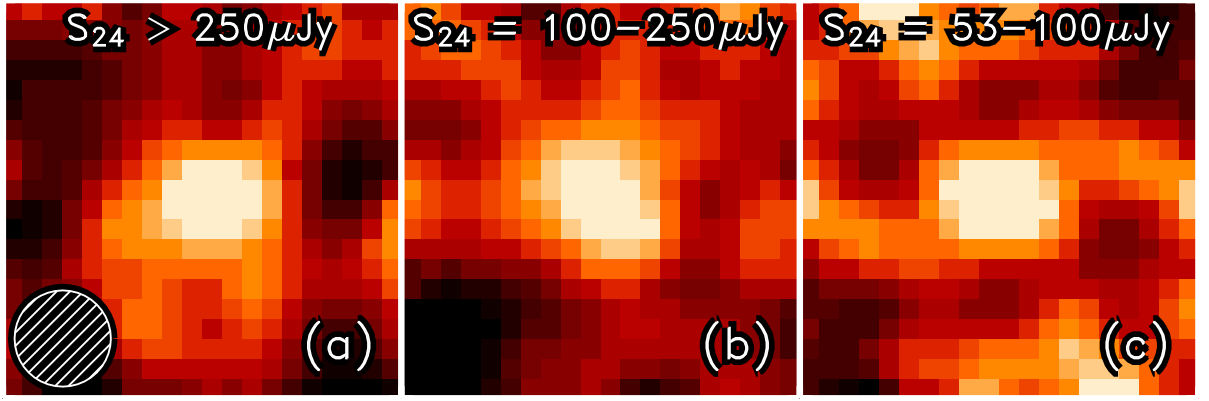}
\caption{Stacked 160~\micron\ images of ordinary IR sources
at $1.5 < \zph < 2.5$ (i.e., those sources without X-ray detections
and IR--power-law-like colors).  The panels show the 160~\micron\
emission in bins of 24~\micron\ flux densities of (a) $S_{24} >
250$~\ujy, (b) $100 \leq S_{24}/\ujy < 250$, (c) $53 < S_{24}/\ujy <
100$.  Each panel shows a region approximately 170\arcsec\ $\times$
170\arcsec\ (roughly 5$\times$ the 160~\micron\ PSF FWHM), at a scale
of 8\arcsec~pix$^{-1}$. The filled circle, inset in panel (a), has a
diameter equal to the PSF FWHM.\label{fig:stackedimage160}}
\end{figure}

Even the deep X-ray data used here is insensitive to heavily obscured
AGN at $1.5 < z < 2.5$, although such objects contribute to
24~\micron--selected samples.  We therefore  also considered a
population of 24~\micron\ sources in this redshift range with red
rest--frame near--IR colors, which is  indicative of the emission from
warm dust at $\lambda \gsim 2$~\micron\ heated by an obscured AGN
\citep[see,
\eg,][]{lac04,stern05,alo06,don07}.   Following \citet{alo06}, we fit
for the exponent in $f_\nu \sim \nu^{\alpha}$ for each 24~\micron\
source with $1.5 < \zph < 2.5$ over the IRAC bands, and selected
objects with $\alpha < -0.5$ (hereafter we refer to this as the
``IR--power-law'' subsample).   We subsequently inspected the SEDs of
all the objects in this subsample visually to ensure that their IRAC
colors are consistent with red power-law--like SED.  Because
power-law--selected AGN have a high 24~\micron--detection
rate \citep[\eg,][]{don07}, the 24~\micron\ selection of the
IR--power-law sample increases the likelihood that they harbor AGN.
The IR--power-law sample includes 76 galaxies (including 13 X-ray
sources and 63 non--X-ray sources) with $S_{24} > 53$~\ujy.
There is a large overlap between the subsample of X-ray sources and
the IR--power-law sources --- 22/45 of the X-ray sources satisfy the
IR--power-law subsample.  These two samples are not independent.

\begin{deluxetable}{lcccccc}
\tablewidth{0pt}
\tablecaption{Average MIPS Flux Densities for $1.5 < z< 2.5$ Galaxies\label{table:stack}}
\tablehead{ \colhead{$S_{24}$ Range} & \colhead{} &
\colhead{$\langle S_{24} \rangle$} &
\colhead{$\langle S_{70} \rangle$} &
\colhead{$\delta(\langle S_{70} \rangle)$} & 
\colhead{$\langle S_{160} \rangle$} &
\colhead{$\delta(\langle S_{160} \rangle)$} \\
\colhead{(\ujy)} & \colhead{N} & \colhead{(\ujy)} & \colhead{(mJy)} & 
\colhead{(mJy)} & \colhead{(mJy)} & \colhead{(mJy)} \\
\colhead{(1)} & \colhead{(2)} & \colhead{(3)} & \colhead{(4)} &
\colhead{(5)}  & \colhead{(6)}  & \colhead{(7)} }
\startdata
\multicolumn{7}{c}{Ordinary IR Sources} \\
$\geq 250$~\uJy  & \phn48 & 380 & 2.0\phn & 0.5\phn & 4.6 & 2.6 \\
$[100,\,250)$~\ujy & 149 & 160 & 1.5\phn & 0.3\phn & 5.1 & 2.2  \\
$[53,\,100)$~\ujy &   \phn77 & \phn95 & 0.88 & 0.35 & 4.6 & 2.6 \\
[4pt] \hline
\multicolumn{7}{c}{X-ray Sources} \\
$\geq 250$~\uJy     &  \phn17 & 480 & 2.2\phn & 0.7\phn & 7.1 & 4.3 \\
$[53,250)$~\ujy & \phn28 & 150 & 0.67 & 0.50 & 3.4 & 3.0 \\
[4pt] \hline
\multicolumn{7}{c}{IR power-law Sources} \\
$\geq 250$~\ujy & \phn24 & 530 & 4.0\phn & 0.7\phn & 10.5 & 4.3 \\
$[53,250)$~\ujy & \phn52 & 170  & 1.4\phn & 0.5\phn & 3.8 & 2.7
\enddata
\end{deluxetable}

For each sub--sample of galaxies, we stacked the locations of each
24~\micron\ source in the 70 and 160~\micron\ images.  We then
measured the flux density in the mean image constructed from the
stack.   Figure~\ref{fig:stackedimage} shows the mean of the stacked
70~\micron\ images for the ordinary IR galaxies split into
subsamples of 24~\micron\ flux density.
Figure~\ref{fig:stackedimage160} shows the stacked 160~\micron\ images
for these galaxies.  Table~\ref{table:stack} gives the measured flux
densities for each of the sub--samples.  

In all subsamples of ordinary IR galaxies, we detect the
source at 70 and 160~\micron\ in the stacked images.  We detect the
X-ray and IR--power-law sources in the 70 and 160~\micron\ images for
the $S_{24} > 250$~\ujy\ subsamples.  However, we do not detect these
subsamples at 70 and 160~\micron\ in the lower flux--density bins, $53
< S_{24}/\ujy < 100$ nor $100 < S_{24}/\ujy < 250$, primarily owing
to the lack of objects in these subsamples.   We do detect the 70 and
160~\micron\ emission  for X-ray sources and IR--power-law sources
within a larger bin, $53 \leq S_{24} < 250$~\ujy, and we quote these values in
table~\ref{table:stack}.  

While the profile of the source in all the stacked images is
consistent with the PSF and 70 and 160~\micron, we observe some extra
power in the wings of the stacked sources.   We suspect this arises from
slight astrometric offsets between the sources at 24~\micron\ and the
sources in the 70 and 160~\micron\ images that cause a ``blurring'' of
the source in the 70 and 160~\micron\ stacked images.   While there is
no way to correct for this, we estimate that it reduces the derived
flux densities by $\lsim$10\% based on differences between the
observed profile and theoretical PSFs.

The choice of the flux--density ranges used to delimit the subsamples
in Table~\ref{table:stack} is pragmatic, and provides sufficient
numbers of objects in each bin, while minimizing the uncertainties on
the stacked measurement.  We have tested how changes in our
flux--density limit affect the stacked measurement.   Changing the
limit of the highest flux density bin from $>250$ to $>500$~\uJy\ (or
values between) has no affect on the measured $S_{70}/S_{24}$ from the
stacking, although we find that the uncertainty on the stacked value
increases as the number of sources decreases in the higher
flux--density subamples.  We do observe an increase in the stacked
$S_{70}$ measurement as we decrease limit of the highest flux density
bin to values less than 250~\uJy.   Therefore, our choice of
250~\uJy\ as the limit for the high flux density bin provides the
logical choice to study the average $S_{70}$ flux density as a
function of $S_{24}$.

\begin{figure}[tp]
\epsscale{1.0}
\plotone{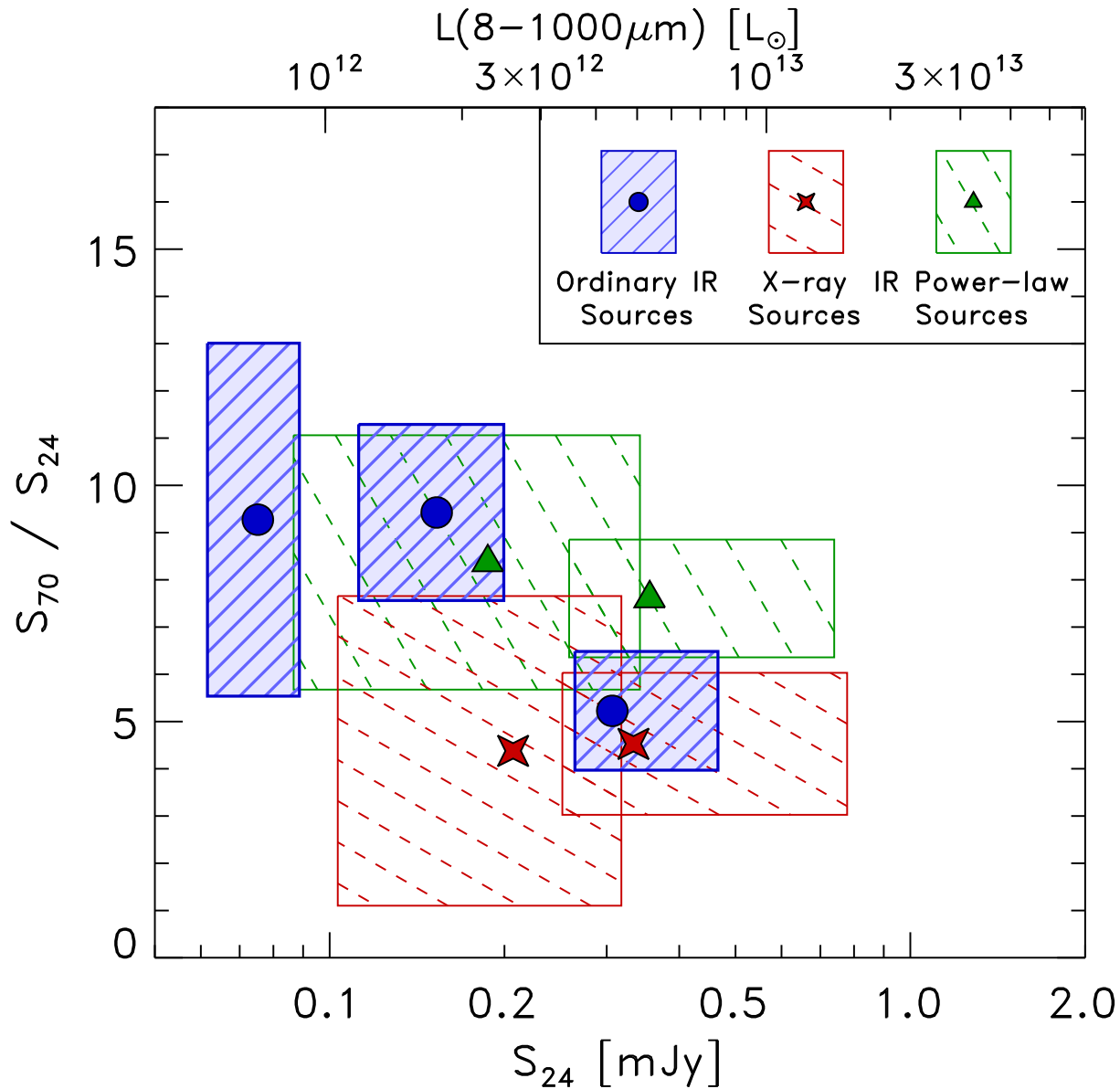}
\epsscale{1.0}
\caption{Measured $S_{70}/S_{24}$ ratios of galaxies at $1.5 < z < 2.5$ derived from
stacking 24~\micron--detected sources in the 70~\micron\ image.  The
symbols show the measured $S_{70}/S_{24}$ ratios from the stacked
images for the ordinary IR sources (without X-ray detections of
IR--power-law-like colors; blue circles), X-ray
sources (red stars), and IR--power-law sources (green triangles), as a
function of 24~\micron\ flux density.   The horizontal width of each
box shows the inter--68\%--tile of the 24~\micron\ flux density
distribution for the sources in each subsample.   The vertical size of
a box shows the 68\% confidence interval on the average flux density
given the measured value and the number of sources in the stack (see
text).  The upper axis shows the total \lir, corresponding to the
measured 24~\micron\ flux density at $z=2$ and using bolometric
corrections from Dale \& Helou (2002) IR templates. \label{fig:f70f24}}
\end{figure}

Figure~\ref{fig:f70f24} shows the average $S_{70}/S_{24}$ ratios for
the galaxies at $1.5 < z < 2.5$ split into subsamples listed in
Table~\ref{table:stack}.    Because the average values at 160~\micron\
have lower S/N, we do not display them graphically.  For ordinary IR sources, the average 70~\micron\ flux densities range from 0.88 to
2.0~mJy over the range of 24~\micron\ flux densities.    These average
values are consistent with those reported by \citet{dye07} using a
smaller sample of galaxies at these redshifts.  The  $S_{70}/S_{24}$
ratios are roughly comparable for the $53 < S_{24}/\ujy < 100$ and
$100 < S_{24}/\ujy < 250$ samples with  an average $S_{70}/S_{24}$
$\approx$9.  However, ordinary IR sources with $S_{24} > 250$~\ujy\
have a $S_{70}/S_{24}$ ratio of $\approx$5, significantly lower than
that measured for fainter $S_{24}$ sources (although within the
uncertainty for the faintest 24~\micron\ subsample).

The $S_{70}/S_{24}$ ratios for X-ray sources show little dependence on
the 24~\micron\ flux density.   Although the range of $S_{70}/S_{24}$
ratios for the X-ray sources with $53 < S_{24}/\ujy < 250$ is large
owing to the uncertainties, these X-ray sources have lower
$S_{70}/S_{24}$ ratios than the ordinary IR sources of comparable
24~\micron\ flux density.  Interestingly, the X-ray and ordinary IR sources with
$S_{24} > 250$~\ujy\ have comparable $S_{70}/S_{24}$ ratios, possibly
implying a similar emission source for the IR emission.

The IR power-law sources with $53 < S_{24}/\ujy < 250$ have
$S_{70}/S_{24}$ ratios generally larger than the X-ray sources of
comparable 24~\micron\ flux density, and are consistent with the
$S_{70}/S_{24}$ ratios of the ordinary IR galaxies.  However,
the IR--power-law galaxies with $S_{24} > 250$~\ujy\ have
$S_{70}/S_{24} \approx 8$, larger than the X-ray and ordinary IR subsamples.  

\section{Discussion}\label{section:discussion}

In this section we discuss the ramification of the average 70 and
160~\micron\ flux density ratios have for the interpretation of the IR
emission from distant galaxies.   Because we detect the average
160~\micron\ images with low significance, we focus the discussion on
the more robust $S_{70}/S_{24}$ values.  However, we use the
160~\micron\ flux densities to constrain the IR luminosities of
distant galaxies in \S~4.3.   We begin with a discussion of the
expected $S_{70}/S_{24}$ ratios of galaxies at $1.5 < z < 2.5$ for
theoretical models and empirical templates for the IR emission of
galaxies.  We then use models and templates to constrain the total IR
luminosity from star formation and AGN in high redshift galaxies.

\subsection{On the Interpretation of $S_{70}/S_{24}$ Ratios of $z=2$ Galaxies}\label{section:ratios}

There are many suites of  empirical models and templates describing
the IR SEDs of galaxies
\citep[\eg,][]{ sil98, dev99, row01,cha01,dal02,lag03,sie06}.   Here,
we use primarly two sets, the empirical model templates of \citet[DH02
hereafter]{dal02} and the theoretical models of \citet[SK07
hereafter]{sie06}, to study how the 24 and 70~\micron\ flux densities
of $1.5 < z < 2.5$ galaxies relates to the total IR luminosity.   SK07
generated theoretical IR SEDs for stellar populations and AGN embedded
in clouds of dust and gas for a wide range of model ionization
luminosity, cloud size and density, and extinction.   SK07 showed that
combinations of these model parameters  fit the IR SEDs of known local
IR--luminous galaxies.  

DH02 parameterized their models by heating intensity.   Following
\citet{lag03}, we use an empirical model, which assigns each DH02
template to a given \lir\ for its \iras\ $S_{60}/S_{100}$ color using
the empirical equation of \citet{marci06}.   Lagache \etal\ showed
that this broadly matches local \iras\ and \iras--sub-mm color
distributions as a function of \lir, although this does not allow for
dispersion in the IR colors \citep[\eg,][]{cha03}.   \citet{marci06}
presented evidence that these models provide reasonable agreement
between total IR luminosities derived from the rest--frame
8--15~\micron\ and 15-24~\micron\ emission, and from the radio
emission for galaxies with $\lir \lsim 10^{12}$~\lsol\ and $z\lsim
1$, supporting the assertion that these models represent the mid--IR
properties of high redshift galaxies.

\begin{figure}[tp]
\epsscale{1.14}
\plottwo{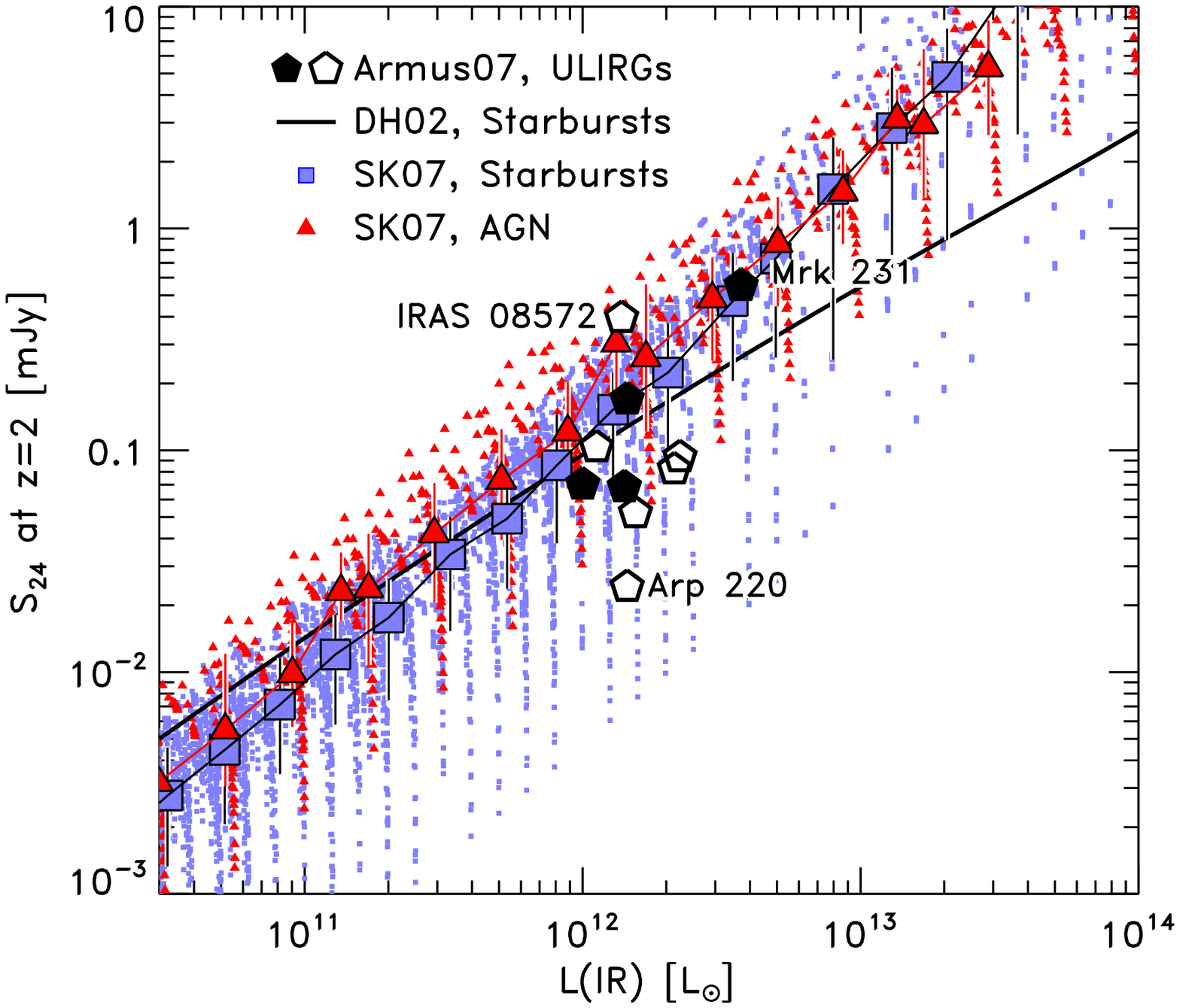}{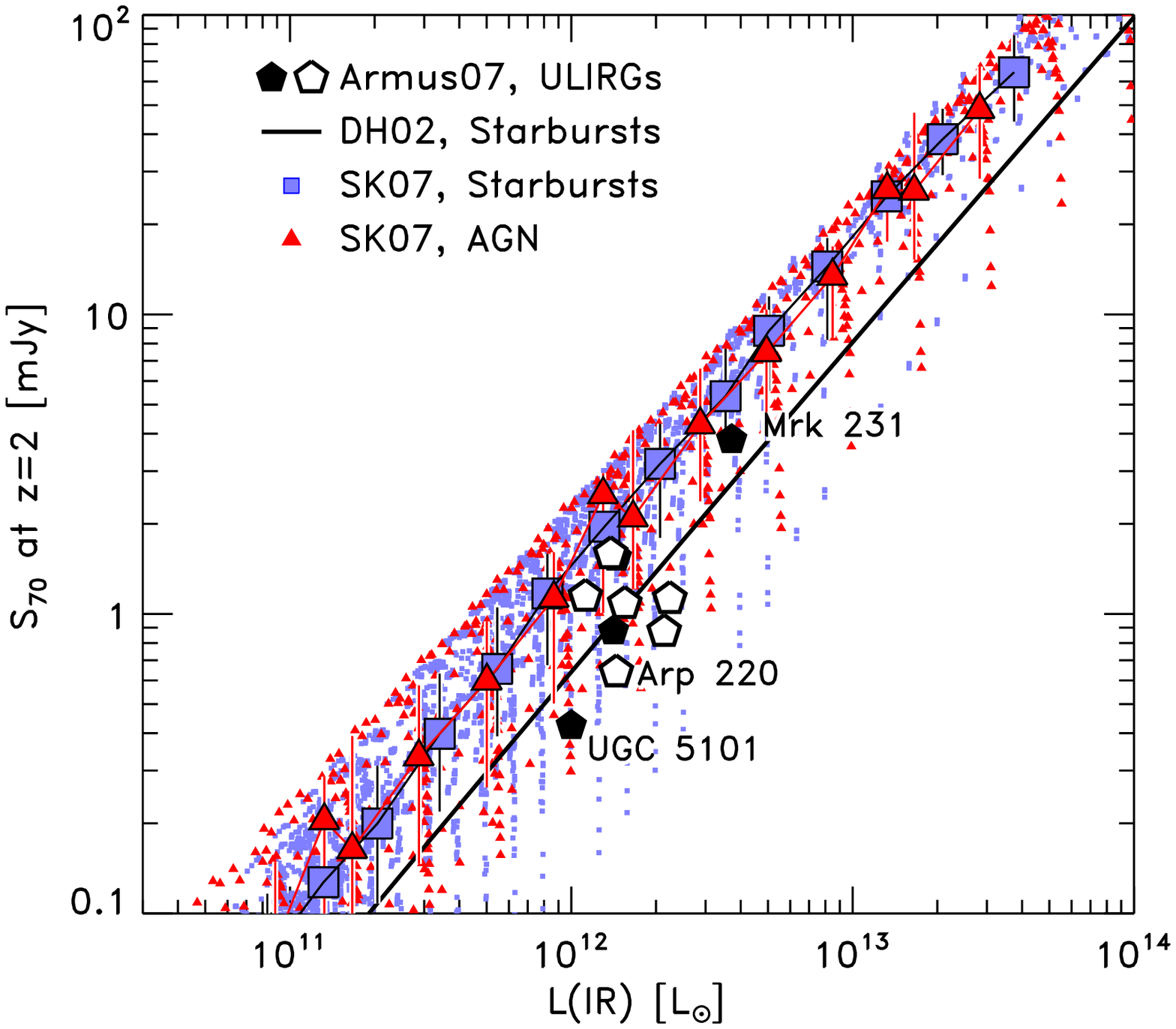}
\epsscale{1.0}
\caption{
Predicted MIPS flux densities for galaxy models observed at $z=2$.
The left panel shows the 24~\micron\ flux densities and the right
panel shows the 70~\micron\ flux densities as a function of total IR
luminosity, $\lir \equiv L(8-1000\micron)$.  In each panel the thick,
solid line shows the prediction for our fiducial  model of Dale \&
Helou (2002; DH02).  Small, filled boxes and triangles show
predictions for the star-forming and AGN theoretical models  of
Siebenmorgen \& Kr\"ugel (2006; SK07), respectively.  Large boxes and
triangles with solid vertical lines show the median and
inter--68\%--tile range of the SK07 models.  The large pentagons show
the expected MIPS flux densities for the IRAS-selected ULIRGs of Armus
et al.\ (2007) if observed at $z=2$.  Filled pentagons show those
ULIRGs with high X-ray to far-IR luminosities.  ULIRGs with extreme
flux densities are labeled.
\label{fig:f24f70atz2}}
\end{figure}

\begin{figure}[tp]
\epsscale{1.0}
\plotone{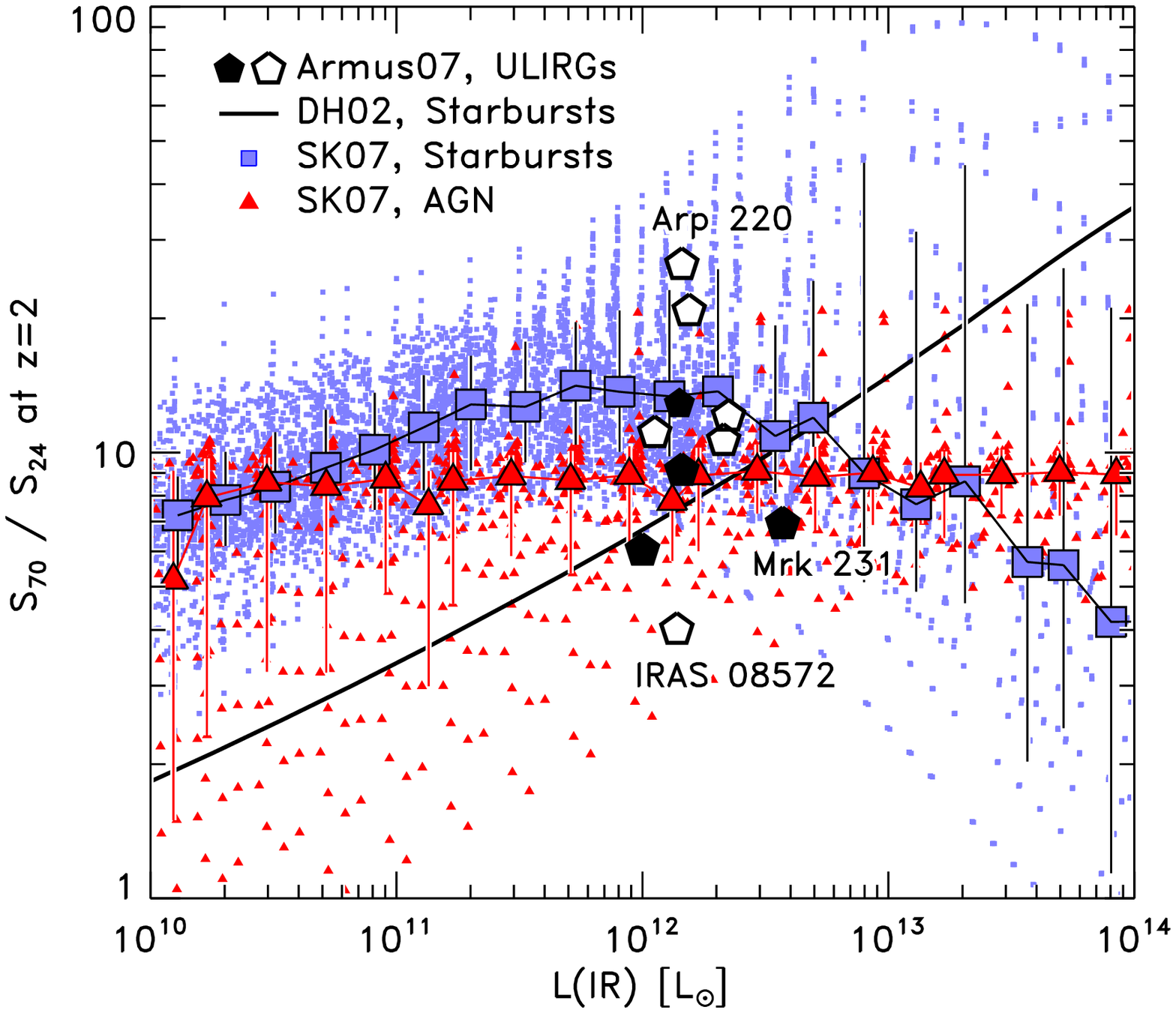}
\epsscale{1.0}
\caption{
Predicted $S_{70}/S_{24}$ ratio at $z=2$ as a function of \lir\ for
the  DH02 and SK07 models.   As in figure~\ref{fig:f24f70atz2} the
thick, solid line shows the prediction for the empirical DH02 models,
and the filled squares and triangles show the SK07 theoretical models
for star-forming galaxies and AGN, respectively.  Large, filled
symbols and vertical lines show the median and inter--68\%--tile range
for the SK07 models.  The large pentagons show the expected
$S_{70}/S_{24}$ ratios for the IRAS-selected ULIRGs of Armus et al.\
(2007) if observed at $z=2$.  Filled pentagons show those ULIRGs with
high X-ray to far-IR luminosities.  ULIRGs with the extreme ratios are
labeled.
\label{fig:f70tof24atz2}}
\end{figure}

Figure~\ref{fig:f24f70atz2} shows the predicted 24 and 70~\micron\
flux densities for the DH02 and SK07 as observed at $z=2$ (rest--frame
8 and 24~\micron) as a function of model
\lir.   The 24 and 70~\micron\ flux densities for $z=2$ galaxies
predicted by the DH02 model increase monotonically with \lir.   The
SK07 models predict a large range of flux density for a given \lir,
spanning approximately an order of magnitude in $S_{24}$ and
$\sim$0.5~dex in $S_{70}$.  For the SK07 models, the
median value and inter--68\%-tile range of flux densities for a given
\lir\ generally increases in a similar fashion to the DH02 models
(although these median values have no physical meaning because any
model is as physical as the next).   Interestingly, the DH02 and SK07
models predict different \lir\ for $z=2$ sources with bright $S_{24}$.
A $S_{24} = 1$~mJy source at $z=2$ would have $\lir \approx 3\times
10^{13}$~\lsol\ with the DH02 model, compared to $\lir \approx 6\times
10^{12}$~\lsol\ for the median SK07 model (although possible SK07
models encompass the DH02 model value for any $\lir$).

Observations of local IRAS--selected ULIRGs would have observed--frame
24 and 70~\micron\ flux densities at $z=2$ generally consistent with
the models, but with appreciable scatter.  Figure~\ref{fig:f24f70atz2}
shows the expected 24 and 70~\micron\ flux densities for ULIRGs from
\citet{arm06} at $z=2$.   The expected $S_{70}$ and
\lir\ for the ULIRGs are within a factor of order 2 of the DH02
models.   However, while many of the ULIRGs of Armus et al.\ have $S_{24}$ and \lir\
within a factor $\sim$2 of DH02 model, there are outliers that span a large range in
$S_{24}$, from $\approx$25~\ujy\ for Arp~220 to $\approx$400~\ujy\ for
IRAS 08572+3915.

Figure~\ref{fig:f70tof24atz2} shows the $S_{70}/S_{24}$ ratio for the DH02
and SK07 models observed at $z=2$ as a function of \lir.   The
$S_{70}/S_{24}$ ratios of the DH02 models increase monotonically with
\lir, indicating that more luminous sources have higher
$S_{70}/S_{24}$ ratios (although this is in constrast to the
observations, see \S~\ref{section:constraints} below).  The SK07
models span a large range of $S_{70}/S_{24}$ ratio at fixed \lir\
owing to the large range of parameter possibilities.  Interesting
however, all of the SK07 models for star-forming galaxies with $\lir <
10^{12}$~\lsol\ have $S_{70}/S_{24}$ ratios significantly larger than
the DH02 model.   We suspect this is because although the  SK07 models
include many individual OB--star associations, they limit the models to spherical
geometry.   Also, the SK07 models may not include the
superposition of intense star-forming regions with additional IR
emission from extended cold dust in a galactic disk.  The DH02 models
are based on empirical observations of local star forming galaxies,
designed to match IRAS color--luminosity  relations, and include the
complex emission from galaxies.

SK07 models for star-forming galaxies with $\lir \sim
10^{12}-10^{14}$~\lsol\ span two orders of magnitude in
$S_{70}/S_{24}$ ratio. The median $S_{70}/S_{24}$ ratio for these
models declines with increasing \lir, in contrast to our DH02 model.
Thus, the SK07 star-forming--galaxy model predicts that the most luminous
galaxies may have low $S_{70}/S_{24}$ ratios (broadly consistent with
the observations, see below).  In contrast, SK07 models for AGN span the largest
range of $S_{70}/S_{24}$ ratio for less--luminous objects.  The SK07 AGN models
with $\lir > 10^{12}$~\lsol\ span a range of
$S_{70}/S_{24}$ ratios, from $\approx$5--20.

The ULIRGs from \citet{arm06} have flux densities at $z=2$ that span
almost an order of magnitude in $S_{70}/S_{24}$ ratio.  Most of this
scatter results from the range of $S_{24}$.  Excluding Arp~220 and
IRAS~2491-1808 with $S_{70}/S_{24}$ ratios $>20$, and IRAS~08572+3915
with a $S_{70}/S_{24}$ ratio $\approx$4, the other ULIRGs have
$S_{70}/S_{24}$ ratios comparable to the DH02 models and the SK07
median range.  The ULIRGs with low hard-X-ray--to--IR-luminosity
ratios are presumably dominated by star formation.  These objects span
the full range of $S_{70}/S_{24}$ ratios suggesting that star-forming
ULIRGs have diverse mid--IR flux ratios, even at comparable total IR
luminosity.  Focusing instead on the ULIRGs with high
hard-X-ray--to--IR-luminosity ratios that presumably have some
contribution from AGN, these objects match the $S_{70}/S_{24}$ flux
ratios of the SK07 AGN models, as well as the DH02 predictions.   The
expected $S_{70}/S_{24}$ ratios from the local ULIRGs do not depend on
the ratio of the X-ray to IR luminosity.  Observations of objects with
a range of hard-X-ray--to--IR-luminosity ratios at other \lir\ could
perhaps better discriminate AGN and star-forming ULIRG samples.

\subsection{Constraints on the Far--IR Spectral Energy Distributions
of High Redshift Galaxies}\label{section:constraints}

\begin{figure}[tp]
\epsscale{0.9}
\plotone{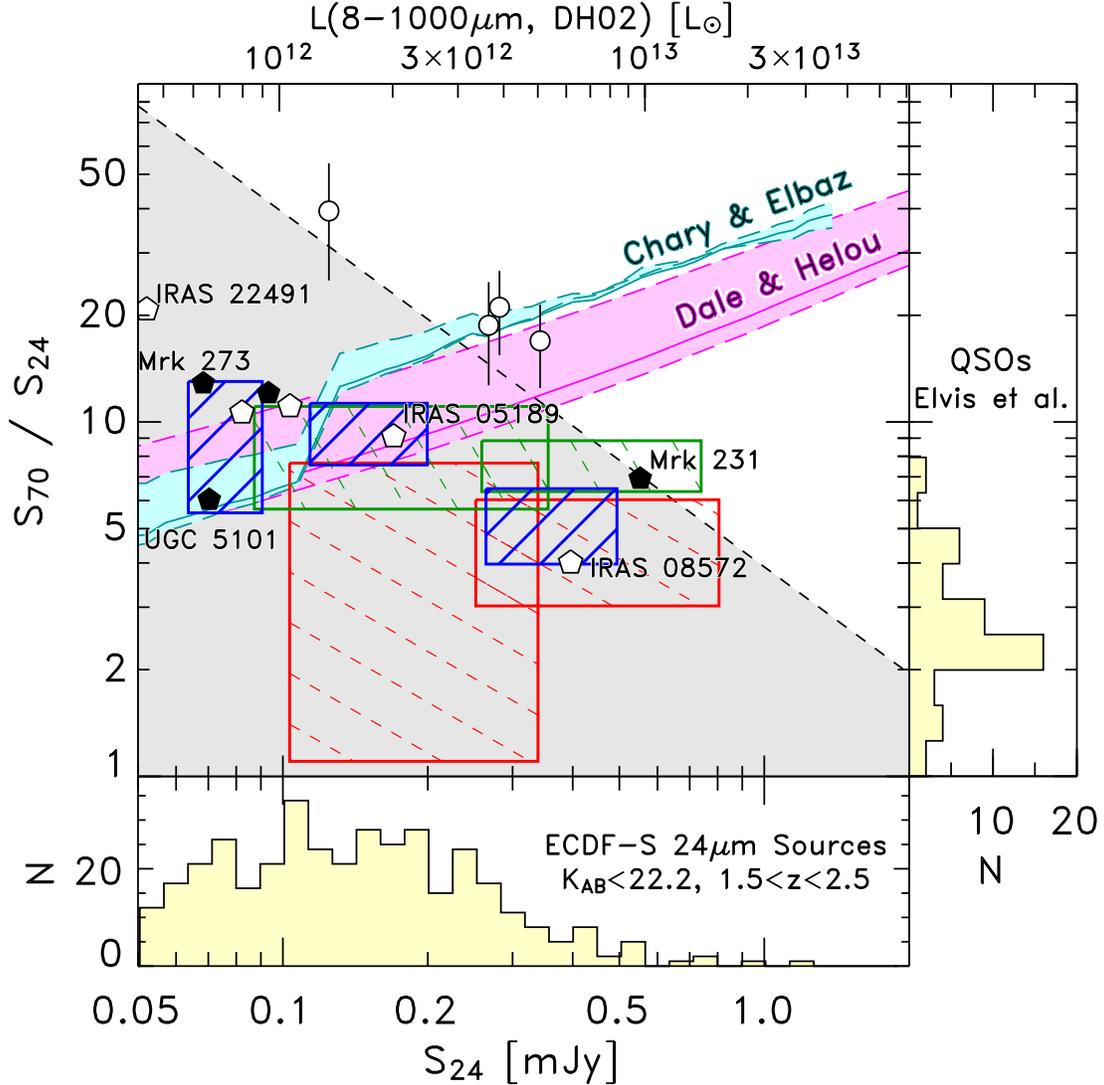}
\epsscale{1.0}
\caption{Comparison of the $S_{70}/S_{24}$ ratios of $1.5 < z < 2.5$
galaxies from the stacking analyses to model expectations.   Hatched
boxes correspond to the average  values with hatch pattern and colors
as in figure~\ref{fig:f70f24}.  Circles show those galaxies with $1.5
< z < 2.5$ detected at 70~\micron.   The pentagons show the expected
$S_{70}/S_{24}$ ratios for the IRAS-selected ULIRGs of Armus et al.\
(2007) if observed at $z=2$.   Filled pentagons show those ULIRGs with
high X-ray to far-IR luminosities.  ULIRGs with the extreme
ratios are labeled.   The magenta--shaded swath shows the expected
flux ratios for the Dale \& Helou (2002) models for  $1.5 < z < 2.5$.
The cyan--shaded swath shows the expected flux ratios for Chary \&
Elbaz (2001) models over the same redshift range.  In both cases the
solid line shows the expected value at $z=2$.  The shaded region
shows the area below the MIPS 70~\micron\
50\% completeness limit (3.9~mJy).    The top abscissa shows the total
\lir\ corresponding to $S_{24}$ at $z=2$ using bolometric corrections
from the Dale \& Helou models. The right panel shows the distribution
of $S_{70}/S_{24}$ ratios for QSO SEDs taken from \citet{elv94}.   A
source with constant power in $\nu f_\nu$ has $S_{70}/S_{24} = 2.9$.
The bottom panel shows the $S_{24}$ distribution of the \ks--band
sample with $1.5 < z < 2.5$.
\label{fig:f70f24_ref}}
\end{figure}

Figure~\ref{fig:f70f24_ref} compares the average $S_{70}/S_{24}$
ratios for the $1.5 < \zph < 2.5$ galaxies from the stacking analysis
to expected ratios for local empirical template SEDs.    The cyan
swath in figure~\ref{fig:f70f24_ref} shows the expected
$S_{70}/S_{24}$ ratios as a function of 24~\micron\ flux density for
the IR model templates of DH02 with $1.5 < z < 2.5$, which include the
\lir\ dependence on dust temperature from \citet{lag03}.  For
comparison, the magenta swath shows the expected $S_{70}/S_{24}$ as a
function of 24~\micron\ flux density from
\citet[CE01 hereafter]{cha01} for the same redshift range.   The
figure also shows the expected $S_{70}/S_{24}$ ratios and 24~\micron\
flux densities at $z=2$ for the local ULIRGs \citep{arm06}, which show
a wide range in diversity in their mid--IR properties as discussed
above.  Here, we discuss the $S_{70}/S_{24}$ ratios for the galaxy
samples at $1.5 < \zph < 2.5$.

\subsubsection{Ordinary IR sources}

The IR luminosities of ordinary IR galaxies at $1.5 < z < 2.5$ (i.e.,
objects without X-ray detections or IR--power-law-like colors) with $53 <
S_{24}/\ujy < 100$ are $10^{11}-10^{12}$~\lsol, using the 24~\micron\
emission with bolometric corrections from either the DH02 or CE01.
The stacked 70~\micron\ flux density for sources at $1.5 < z < 2.5$
with $53 < S_{24}/\ujy < 100$ has S/N$<$3, and some caution must be
applied.  Nevertheless, the measured average $S_{70}/S_{24}$ ratio for
these sources match the values expected from the DH02 and CE01 models.
They are consistent with the theoretical SK07  models, which predict
that star-forming galaxies with  with $53 < S_{24}/\ujy < 100$ and
$\lir \sim 10^{11} - 10^{12}$~\lsol\ have $S_{70}/S_{24}$ $\sim$8--15.
Four of the \citet{arm06} ULIRGs would have $53 < S_{24}/\ujy < 100$
observed at $z=2$.   These sources have $\lir
=1-2\times10^{12}$~\lsol, and $S_{70}/S_{24}$ ratios 6--13, consistent
with the measured values from the stacking.

Sources with $1.5 < z < 2.5$ and $100 < S_{24}/\ujy < 250$ have $\lir
= 1-3\times10^{12}$~\lsol\ using bolometric corrections from DH02 (and
somewhat larger \lir\ using CE01).  The measured $S_{70}/S_{24}$ ratio
from the stacking analysis has S/N=5.   Inspection of
figure~\ref{fig:f70f24_ref} shows that the observed $S_{70}/S_{24}$
ratio is consistent with the DH02 model, but lower by a factor $\sim$2
than expected from the CE01 model.   The measured $S_{70}/S_{24}$
ratio is on the low end of the values  predicted by the theoretical
SK07 models.  The SK07 models that reproduce the measured
$S_{70}/S_{24}$ ratio for these galaxies have high extinction, $A_V =
2.2 - 6.7$~mag (corresponding to $A_{\mathrm{UV}} = 6-19$~mag at
1500~\AA), whereas the inter--68\%--tile range of all SK07 models at
these IR luminosities have even larger extinctions, $A_V = 7 - 37$~mag
(and in comparison, the SK07 model to Arp~220 has $A_V = 72$~mag).

If these models accurately describe the high--redshift
galaxies, then they have high extinction.   However, we note that the
\textit{minimum} extinction in the SK07 models is $A_V = 2.2$,
suggesting that possible models with lower extinction may fit the
measured flux density ratios equally well.  For example,
\citet{dad07a} find that most $z\sim 2$ $BzK$--selected galaxies have
relatively low (``optically thin'') dust attenuations, corresponding to
$A_{\mathrm{UV}} \sim 2-6$~mag.  However, most of the $z\sim 2$ galaxies in
their sample have low 24~\micron\ flux densities, $S_{24} \ll
100$~\uJy.  Furthermore, the $BzK$--selection may miss objects with
high dust obscuration, such as sub-mm galaxies or distant red galaxies
(e.g., see Reddy et al.\ 2006).  As with other observations of
local and distant galaxies, a galaxy's extinction
increases with bolometric luminosity \citep[\eg,][]{wan96,red06}.
Our comparison against the SK07 models implies that the sources with
$100 < S_{24}/\ujy < 250$ at $z=2$ have relatively high extinction as well,
consistent with these other results.

Two of the local ULIRGs from Armus et al.\ (2007) would have $100 <
S_{24}/\ujy < 250$ at $z=2$, IRAS~05189-2524 and IRAS~15250+3609.
They have $\lir \approx 10^{12}$~\lsol, and $S_{70}/S_{24}$ ratios of
$\approx 9$ and 11, comparable to the measured values from the
stacking.  Both of these ULIRGs have emission and absorption features
in their mid--IR spectra that are fairly average in the sample of
Armus et al.\ (2007).  To measure the strength of the mid--IR emission
features directly  for $1.5 < z < 2.5$ galaxies with $100 <
S_{24}/\ujy < 250$ requires prolonged exposures with mid-IR
spectrographs --- the Infrared Spectrograph (IRS) on board \spitzer\ requires $\sim 10$~hrs to obtain spectra of objects with $S_{24} \sim 150-250$~\ujy\ \citep{tep07} ---
or by targeting gravitationally lensed 24~\micron\ sources with IRS
(e.g., J.~Rigby et al., in preparation).   Nevertheless, the few
published IRS spectra of galaxies at these redshifts and 24~\micron\
flux density show that the mid--IR emission features are consistent
with local IR--luminous galaxies.     Therefore there is little
evidence that the $1.5 < z < 2.5$ galaxies with $100 < S_{24}/\ujy <
250$ have abnormally strong or weak mid--IR features.

Ordinary IR galaxies with $S_{24} > 250$~\ujy\ and $1.5 < z <
2.5$ have lower $S_{70}/S_{24}$ ratios compared to the fainter 24~\micron\
sources.  Interestingly, both the DH02 and CE01 models have
$S_{70}/S_{24}$ ratios larger than the measured ratio by factors
$\approx$2--3. This is striking because if there existed a substantial
population of sources with $S_{24} > 250$~\ujy\ and $S_{70}/S_{24}$
ratios high as the DH02 or CE01 model predictions, then these sources
would have $S_{70} \gsim 3-10$~mJy and would have been detected
directly in the 70~\micron\ data.   Of the 89 sources with $S_{24} >
250$~\ujy\ (including 17 with X-ray detections and 24 with IR
power-law SEDs), we detect only three at 70~\micron\ (all with
3$<$S/N$<$4).   Interestingly, these 70~\micron--detected objects have
$S_{70}/S_{24}$ ratios consistent with the DH02  and CE01 models.
Such sources must be extremely rare --- they  represent only $3\pm2$\%
(3/89) of $S_{24} > 250$~\ujy\ sources.  Even more extreme (and rare)
$S_{24} > 750$~\ujy\ and $S_{24} > 900$~\ujy\ sources at $z\sim 2$
studied by \citet{hou05}, and \citet{yan07}, respectively, have
$S_{70}/S_{24}$ flux density ratios $< 10$ (E.~Le~Floc'h, 2006,
private communication; K.~Tyler et al., 2007, in preparation; Sajina et al.\ 2007), even though they were
selected over much larger fields than used here ($\simeq$9 and
$\simeq$4~sq.\ deg, respectively), consistent with our findings.

Therefore, the brightest 24~\micron\ sources at $1.5 < z < 2.5$
have low $S_{70}/S_{24}$ ratios.   These galaxies may suffer from
unusually strong mid--IR emission
features from PAHs, which would boost the observed 24~\micron\ flux
density and lower $S_{70}/S_{24}$.   Several published mid--IR spectra
datasets from \spitzer\ now exist for galaxies with $S_{24} >
250$~\ujy\ and $z\sim 2$ \citep{hou05,yan05,wee06,wee06b}.  While
some do show strong PAH emission, most have rather  featureless spectra.
Therefore, the low $S_{70}/S_{24}$ flux density ratios are likely not
solely a result of unusually strong mid--IR emission features.

The $S_{70}/S_{24}$ flux--density ratios of the $S_{24} > 250$~\ujy\
sources with $1.5 < \zph < 2.5$ are comparable to SK07 theoretical
star-forming models with moderate extinction, $A_V = 2.2$~mag, and
small physical sizes, $R = 0.35$~kpc.   These sizes are smaller  than
the rest--frame optical radii of $z\sim 2$ galaxies
\citep{tru04,pap05}, which is consistent as the IR--emitting region is
likely not larger than the optical size.  A compact IR--emitting
region implies these objects have warmer dust temperatures ($T\sim
40-60$~K) compared to high--redshift sub-mm--selected galaxies, which
require colder dust temperatures ($T\sim 20$~K, e.g., Chapman et al.\
2005) and an extended emission region to reproduce the sub--mm number
counts ($R\gsim$5~kpc; Kaviani et al.\ 2003).   Indeed, Egami et al.\
(2004) and Pope et al.\ (2006) find that the average IR SEDs for
sub--mm sources are consistent with the IR emission from cooler dust
($\sim$30~K).  These lower dust temperatures are similar that of
Arp~220, which would have $S_{70}/S_{24} \approx 30$ if observed at
$z=2$, strongly in contrast to the $S_{24} > 250$~\ujy\ sample
($S_{70}/S_{24} \approx 6$).   Therefore, the nature of the IR
emission in bright 24~\micron--sources at $1.5 < z < 2.5$ is likely
different from that of sub--mm-selected samples.  If star--formation
powers the IR emission in bright 24~\micron\ sources at $1.5 < z <
2.5$, then these sources are likely compact, presumably with higher
dust temperatures
\citep[\eg,][]{chan06}.

Alternatively, the $S_{24} > 250$~\ujy\ sources at $1.5 < \zph < 2.5$
may be powered by AGN or starburst/AGN composites.  Indeed,
\citet{dad07b} present evidence that sources at $z\sim 2$ with anomalously high
24~\micron\ flux densities compared to other star formation
indications host heavily obscured AGN.   As noted previously, the
$S_{70}/S_{24}$ ratios are comparable to those of the X-ray with
$S_{24} > 250$~\ujy.  Given that even deep X-ray data miss a high
fraction of AGN \citep[\eg,][]{don07}, we expect that some portion of
the $S_{24} > 250$~\ujy\ population has an AGN contribution.  In this
case, similar physical emission mechanisms may be at work in all the
$S_{24} > 250$~\ujy\ samples.

\subsubsection{X-ray sources}

The measured $S_{70}/S_{24}$ ratios for X-ray sources at $1.5 < z <
2.5$ span the empirical range observed for QSOs \citep{elv94}.  X-ray
sources with $S_{24} > 250$~\ujy\ have $S_{70}/S_{24}$ ratios on the
red--tail of the QSO distribution.  This  is probably a result that
our sample contains 24~\micron--bright X-ray sources, which presumably
have IR colors similar to redder, optically--selected QSOs.  The
stacked $S_{70}/S_{24}$ ratio for X-ray sources with $53 < S_{24} <
250$~\ujy\  has low S/N.  It may be that the IR emission in these
sources stems both from AGN and star--formation processes (as in the
case of Mrk~231, e.g., Armus et al.\ 2007).   In this case, an AGN
dominates the mid-IR SED in the X-ray sources, heating dust to warmer
temperatures, which boosts the observed--frame 24~\micron\ emission
and lowers the $S_{70}/S_{24}$ ratios compared to what is typically
observed in starbursts.   Both \citet{alex05} and \citet{fra03} show
evidence that sub-mm--selected galaxies at $z\gsim 2$ show evidence
for simultaneous AGN and star--formation activity, where dust heated
by the starburst dominates the sub-mm emission.     To test for
evidence of colder dust in the X-ray--detected galaxies requires flux
density measurements at longer wavelengths, i.e., at rest--frame
$\lambda > 100$~\micron\ (see, e.g., Le~Floc'h \etal\ 2007).

\subsubsection{IR power-law sources}

The IR power-law sources with $1.5 < z < 2.5$ have similar
$S_{70}/S_{24}$ ratios for the $100 < S_{24}/\ujy < 250$ and $S_{24} >
250$~\ujy\ sample, although the $S_{70}/S_{24}$ ratio for IR power-law
sources with $53 < S_{24}/\ujy < 250$ has S/N$\approx$3.
Nevertheless, IR power-law source have low $S_{70}/S_{24}$ ratios
compared to either the star-forming empirical or theoretical SED
models given their 24~\micron\ flux densities.   SK07 AGN models that
match the $S_{70}/S_{24}$ ratios of IR--power-law sources with $53 <
S_{24}/\ujy < 250$ span a wide range of model parameter space,
providing few useful constraints.    The SK07 models for AGN with
$S_{24} > 250$~\ujy\ that match the IR power-law source
$S_{70}/S_{24}$ ratios span a wide range of extinction, $A_V =
1-128$~mag.   Most of the the models require obscuring regions with
sizes 8--16~kpc.   Therefore, if solely an AGN is responsible for the
IR emission, the extinction region may extend to the entire host
galaxy.  Alternatively, a more plausible scenario is that the IR
emission in these galaxies results from a composite starburst/AGN.

\subsection{Implications for the Total IR Luminosities of
High--Redshift Galaxies}\label{section:irimplications}

The average 24, 70, 160~\micron\ flux densities constrain the shape
and normalization of the average far--IR SED --- and thus the total IR
luminosity --- of the typical 24~\micron\ source at $1.5 < z < 2.5$.
Because we have only the average flux densities, these results apply
only on \textit{average} for the 24~\micron\ source population at $1.5
< z < 2.5$.  There is likely significant variation from object to object.
Indeed, if the IRAS-selected ULIRGs of \citet{arm06} were observed at
$z=2$, then they would show a large variation between
\lir, $S_{24}$ and $S_{70}$  (see figure~\ref{fig:f70f24_ref}).

We fit the measured average 24, 70, and 160~\micron\ flux densities
data in table~\ref{table:stack} for the ordinary IR sources, X-ray
sources, and IR--power-law sources with the DH02 and SK07 model
templates for model redshifts $z=1.5$, 2.0, and 2.5.    Including the
analysis over the full range of redshifts, as opposed to restricting
it to only the average redshift of the sample, shows how the
interpretation depends as a function of redshift.  For example,
because the strong spectral features move through the 24um band at
these redshifts, the redshift of the galaxy makes a sizeable
difference in the interpretation of its $S_{70}/S_{24}$ measurement.
However, the the fits at the three different redshifts are not
statistically independent.   Here, we fit the full range of DH02
template SEDs, with the model normalization as a free parameter.
This allows the high--redshift galaxies to have any ionization for a
given \lir.   For each set of average 24, 70, and 160~\micron\ flux
densities listed table~\ref{table:stack}, we take the IR--luminosity
of the best--fitting DH02 model, scaled to match the measured flux
densities.   Because the SK07 models are derived for a specified set
of physical input parameters, allowing variations in the normalization
would change the physical conditions of the model.   Therefore, we do
not allow the normalization to vary.   Instead, we take the IR
luminosity of the best-fitting SK07 model as the IR luminosity for the
set of average 24, 70, and 160~\micron\ flux densities.

\begin{figure}[tp]
\epsscale{1.0}
\plotone{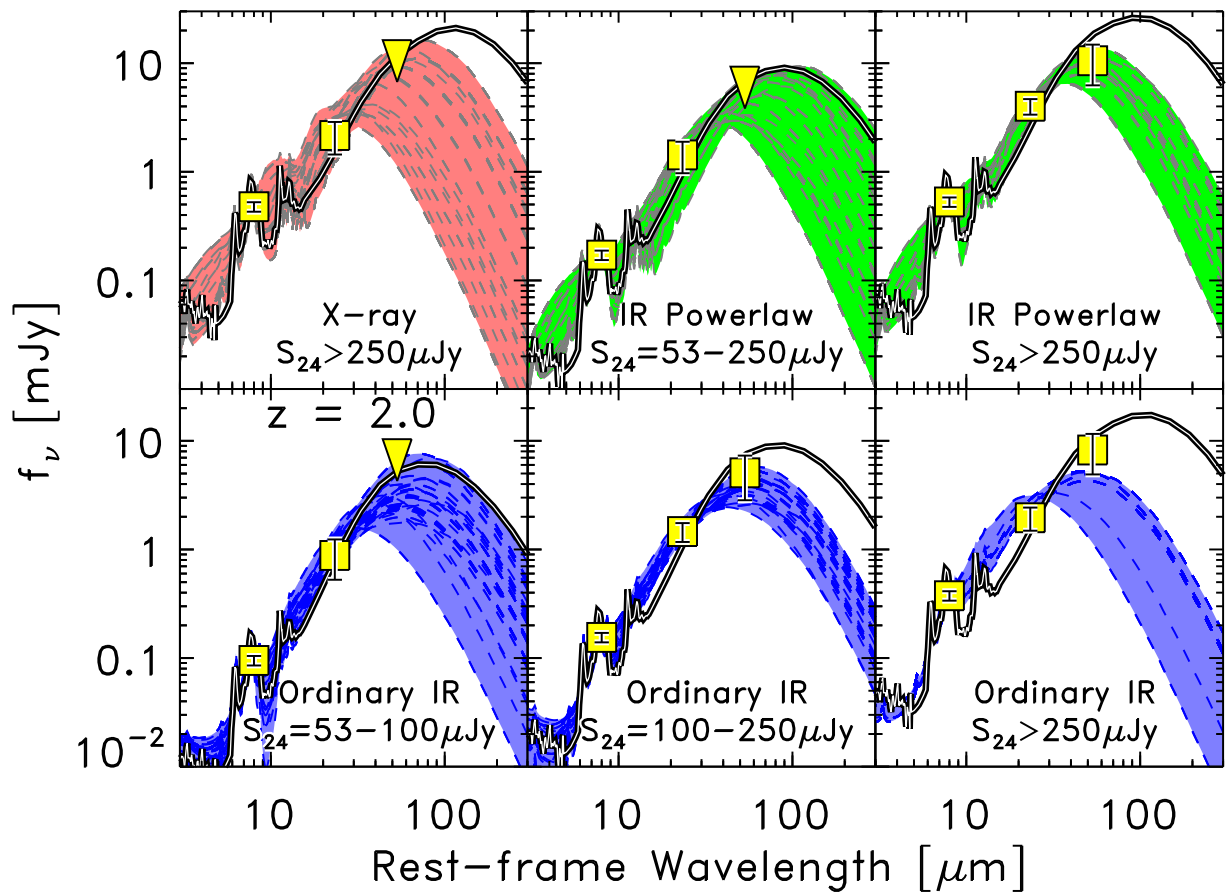}
\epsscale{1.0}
\caption{Model fits for $z=2$ to the 24, 70, and 160~\micron\ flux
densities data listed in  table~\ref{table:stack}.  The yellow squares
in each panel show the measured flux densities.  Downward triangles
show the $1\sigma$ upper limit for measured flux densities in
table~\ref{table:stack} with S/N$<$2.  In each panel, the thick line
shows  the best--fit DH02 model to the data, where the normalization
is a  free parameter.  The shaded area shows the range of SK07 models
that fit the data (without allowing for a normalization of the model
luminosities).  The IR--power-law sources show the results for $53 <
S_{24}/\ujy < 250$~\ujy\ and $S_{24} > 250$~\ujy.  The X-ray sources
show the results for $S_{24} > 250$~\ujy\ only because X-ray sources
with lower 24~\micron\ flux density have S/N$<$2 for both the
70~\micron\ and 160~\micron\ stacked measurements.  For both the X-ray
and IR--power-law sources we use the SK07 AGN models for the fit.  We
fit the SK07 models to the data for the ordinary IR sources.
\label{fig:modelfit}}
\end{figure}

Figure~\ref{fig:modelfit} illustrates the best fitting models to the
ordinary IR sources, X-ray sources, and IR--power-law sources assuming
the  fiducial redshift $z=2$.   In each panel, the thick, solid line
shows the best--fit DH02 model.   We show the range of SK07 models fit
to the data as the shaded region, where we have repeated the model
fitting to the data after randomly adjusting the measured flux
densities by their $1\sigma$ uncertainties.    Although the average
160~\micron\ flux densities have low S/N, they improve the constraints
on the derived IR luminosity, especially by limiting the range of the
SK07 model parameter space.   Nevertheless, in many cases a  fairly
wide range of models fit the data, giving a spread in the inferred
\lir.   While the average 24, 70, and 160~\micron\ data improve the
constraints on the total IR luminosity, significant uncertainties
persist.  Observations at rest--frame wavelengths $>100$~\micron\ are
required to measure the full shape of the IR SED.

Because nearly all studies of high--redshift MIPS--detected galaxies
rely on 24~\micron\ observations only, we also compute the total IR
luminosity using the 24~\micron\ flux density and our fiducial DH02
model (see above; e.g., Le~Floc'h \etal\ 2005, Papovich \etal\ 2006).
While the DH02 models are inappropriate for AGN, most work in the
literature uses similar SEDs to convert the 24~\micron\ flux density
to IR luminosity.   Therefore, it is prudent for us to compare against
them here.    We do not infer the \lir\ from the 24~\micron\ flux
density using the SK07 models as these models span such a large range
of \lir\ for a given $S_{24}$ at these redshifts (see
figure~\ref{fig:f24f70atz2}). 

\begin{figure}[th]
\epsscale{1.0}
\plotone{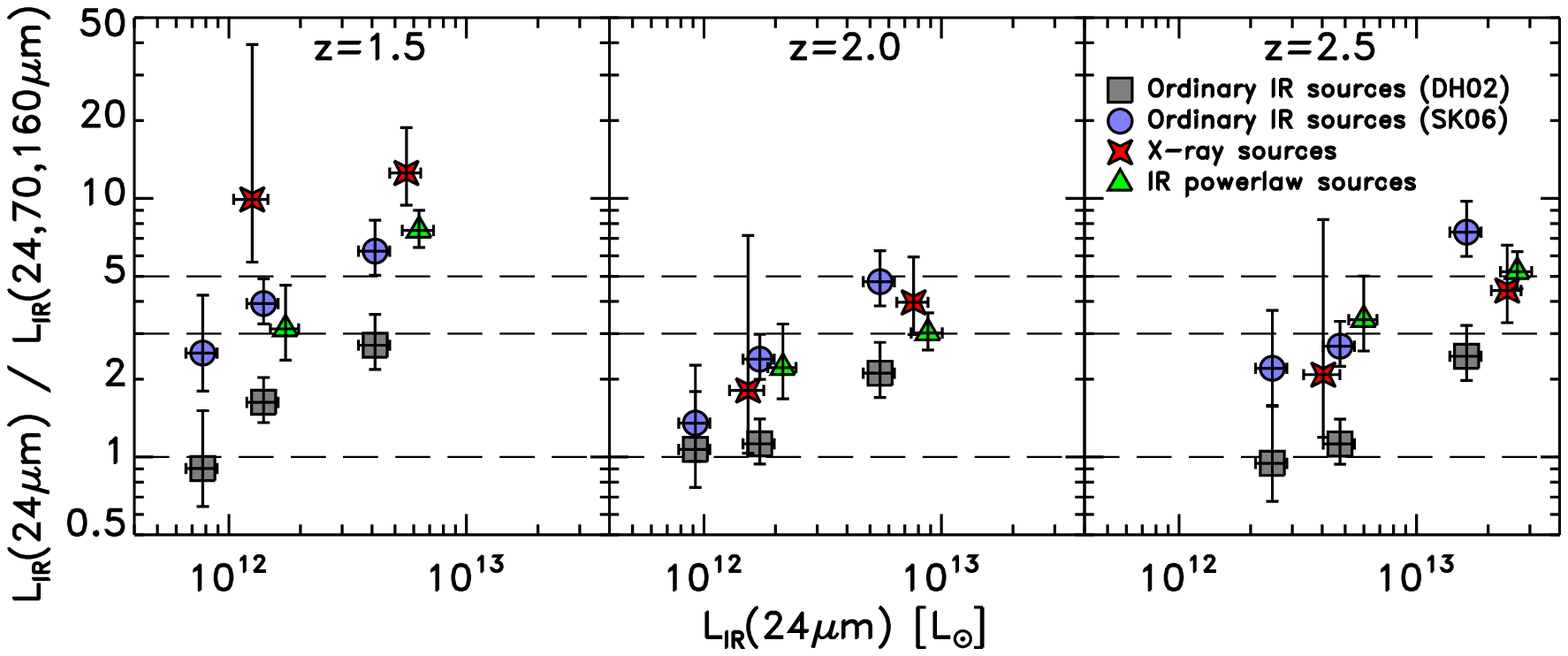}
\epsscale{1.0}
\caption{Comparisons of $\lir(24\micron)$ to $L(24,70,160\micron)$.
The $\lir(24\micron)$ are derived solely using the 24~\micron\ flux
density, $S_{24}$, and specified redshift with the DH02 bolometric
corrections.  The $\lir(24,70,160\micron)$ are derived  using both the
SK07 or  DH02 models for bolometric corrections.  The symbols show the
results for ordinary IR sources (without X-ray detections or
IR--power-law-like colors; blue circles for SK07 models;
gray squares for DH02 models, see plot inset), X-ray sources (red
stars), and IR--power-law sources (green triangles) using the flux
densities listed in table~\ref{table:stack}.   The results for the
ordinary IR sources are shown using both the DH02 and
star-forming SK07 models.   The $\lir(24,70,160\micron)$ values for the
X-ray sources and IR--power-law sources are derived using the SK07 AGN
models.  Each panel shows the fit using models at $z=1.5$, 2, and 2.5,
as labeled.   The horizontal dashed lines show luminosity ratios of 1,
3, and 5.
\label{fig:lir2470vlir24}}
\end{figure}

Figure~\ref{fig:lir2470vlir24} compares the total IR luminosities
derived from the best--fit models to the 24, 70, and 160~\micron\ flux
densities ($\equiv$$\lir(24,70,160\micron)$) to the IR luminosity
estimated solely from the 24~\micron\ flux density
($\equiv$$\lir(24\micron)$).     For all redshifts $1.5 < z < 2.5$,
the $\lir(24,70,160\micron)$ for ordinary IR sources and
$100 < S_{24}/\ujy < 250$ or $53 < S_{24}/\ujy < 100$ are within a
factor of $<$3 of $\lir(24\micron)$.  This is true in the comparison
between $\lir(24\micron)$ and $\lir(24,70,160\micron)$ for
either the DH02 or SK07 models.

The ordinary IR sources with $S_{24} > 250$~\ujy\ generally
have larger $\lir(24\micron)$ / $\lir(24,70,160\micron)$ than fainter
24~\micron\ sources.    The $\lir(24,70,160\micron)$ derived from
the DH02 models are within a factor of $\lsim$3 of $\lir(24\micron)$.
In contrast $\lir(24,70,160\micron)$ derived with the SK07 models
are $\approx$4--8$\times$ lower than $\lir(24\micron)$.  

The $\lir(24,70,160\micron)$ derived with the DH02 models is always a
factor $\approx$2 lower than $\lir(24,70,160\micron)$ derived from the
SK07 models.   This results from the intrinsic difference between the
shape of the SED in the theoretical and empirical models.  As
discussed above, this difference results because in the theoretical
models the IR emission comes from embedded OB--star associations.
Compared to the empirical model, which includes the integrated
emission from complex galactic structures, the theoretical models
require warmer dust temperatures to match the observed flux densities,
with increased relative emission at rest-frame $\lambda \sim
10-30$~\micron\ and decreased relative emission at rest-frame $\lambda
> 50$~\micron\ (see figure~\ref{fig:modelfit}).   The current data do
not support either the IR SED shape from the theoretical or empirical
model, and constraints at rest--frame wavelengths longer than
50~\micron\ are needed.  The different models give a range of
uncertainty arising from the shape of the choice of SED, which
accounts for the factor of two.

While we have not included the analysis of the CE01 models in
figure~\ref{fig:lir2470vlir24}, we make some qualitative comparisons
here.  Given that the CE01 and DH02 models give similar expected
$S_{70}/S_{24}$ ratios for sources with $S_{24} \lsim 100$~\uJy\ at
$z=2$ (figure~\ref{fig:f70f24_ref}), we expect similar results between
these models for the faint sources.   At higher 24~\micron\ flux
densities, the CE01 models predict higher $S_{70}/S_{24}$ ratios, and
thus we expect the difference between $\lir(24\micron)$ and
$\lir(24,70,160\micron)$ to be larger for CE01 than for DH02.

The ratio of $\lir(24\micron)$ to $\lir(24,70,160\micron)$ for the
X-ray and IR--power-law samples depends strongly on the assumed
model redshift and $S_{24}$.   For X-ray and power-law sources with
$S_{24} > 250$~\ujy, this ratio ranges from a factor of $\gsim$10 at
$z=1.5$ to $\sim$3 at $z=2.5$.  The reason for the dramatic decrease
in IR luminosity ratio with increasing redshift is because at $z=1.5$
the SK07 AGN models that fit the data have steeply rising mid-IR SEDs
with very hot dust temperatures, peaking at $\sim$30~\micron.  The
DH02 models place the 24~\micron\ band in the gap between strong PAH
features at 7.7~\micron\ and 11~\micron, with the peak emission at
$\sim 100$~\micron, producing higher IR luminosity for the same
mid--IR flux density.   

\subsection{Implications for the Star Formation of High--Redshift Galaxies }

In this section, we study the implications the stacking results have
for the total IR luminosities and the SFRs for individual
24~\micron--detected galaxies at high redshifts.   To illustrate this,
we compute the total IR luminosities for all the $1.5 < \zph < 2.5$
galaxies with $S_{24} > 53$~\ujy\ using two methods.   We used the
$S_{24}$ and the photometric redshift of each galaxy to infer its
total IR luminosity, $\lir(24\micron)$, using  bolometric corrections
from our fiducial DH02 model,  following the method in \citet{lef05}
and \citet{pap06}.   Second, we assign to each galaxy the 70 and
160~\micron\ flux density derived from the stacking analysis, $\langle
S_{70}\rangle$ and $\langle S_{160} \rangle$ for its measured
$S_{24}$.   We then fit models to the $S_{24}$, $\langle
S_{70}\rangle$, and $\langle S_{160}\rangle$ for each galaxy  at the
redshift of each galaxy to derive $\lir(24,70,160\micron)$, following
the procedure in \S~\ref{section:irimplications}.   For the ordinary
IR sources with $S_{24} > 53$~\ujy\ we used $\lir(24,70,160\micron)$
derived with the DH02 empirical models.  For X-ray and IR power-law
sources with $S_{24} > 53$~\ujy\ we used the SK07 theoretical AGN
models.

Our choice to apply the DH02 models for ordinary IR sources is
motivated by the fact that the theoretical SK07 models do not include
additional dust components from the galaxies' ISM.  Using the SK07
theoretical models would reduce the derived \lir\ by a factor $\approx
2$ (see figure~\ref{fig:lir2470vlir24}).   Although we use the SK07
AGN models for X-ray sources, this may not be fair as these
theoretical models allow for no IR emission due to dust heated by star
formation.  X-ray sources may have additional cold dust components
possibly associated with star--formation \citep[\eg,][]{fra03,alex05}.
If such additional components exist in AGN, then the
\lir\ values derived from the SK07 models may underestimate the total
IR luminosity.\footnote{We note that the SK07 AGN models may not
underestimate the total IR luminosity as there are few extreme SK07 models
where the expected 24~\micron\ flux density at $z=2$ (for a given
\lir) is higher for star-forming regions than for AGN, see
figure~\ref{fig:f24f70atz2}.}

\begin{figure}[tp]
\epsscale{1.0}
\plotone{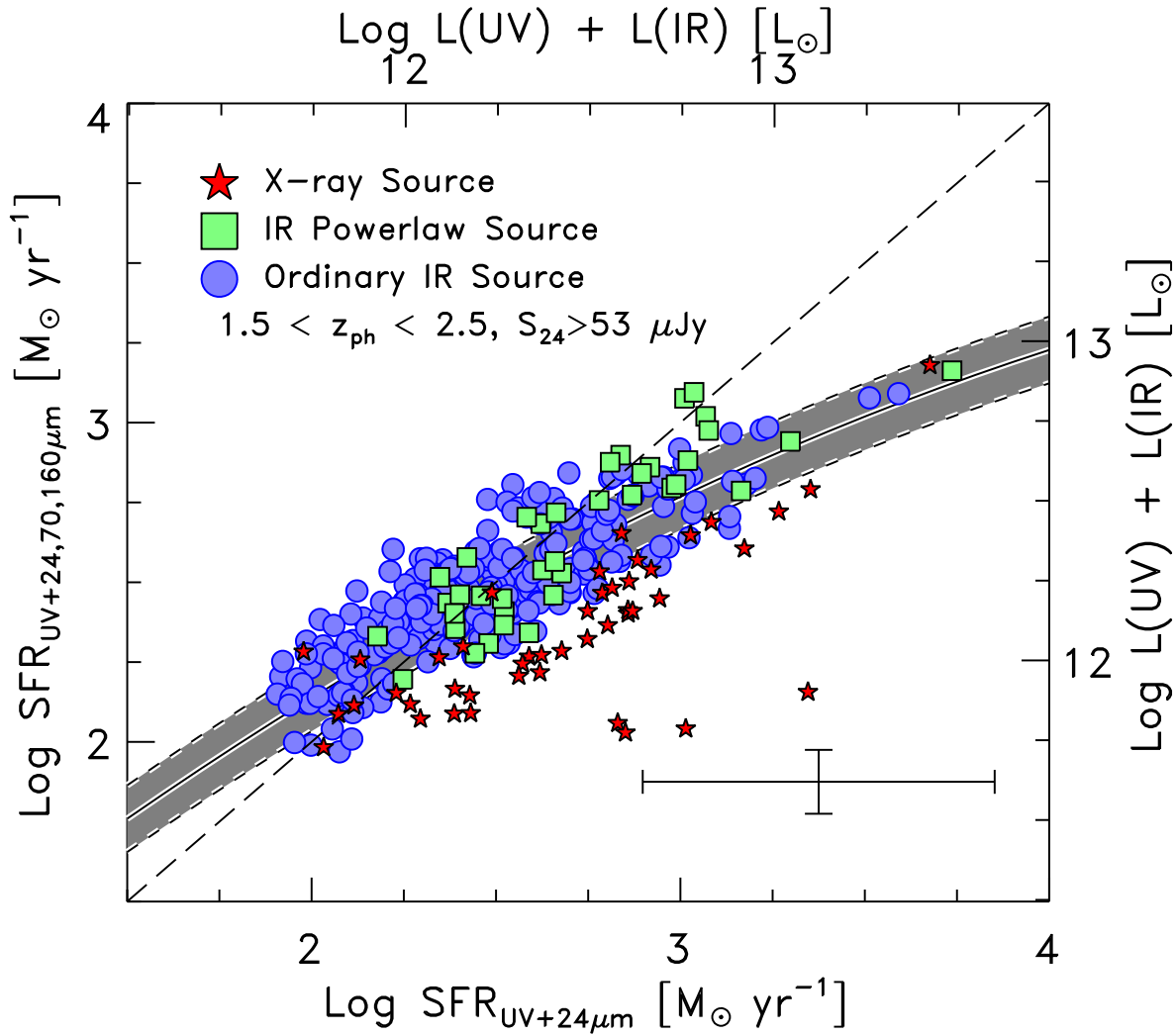}
\epsscale{1.0}
\caption{The SFR derived for galaxies in the MUSYC 
$1.5 < \zph < 2.5$ sample using only the UV and 24~\micron\ data,
SFR$_{\mathrm{UV}+24\micron}$, compared to the SFR derived using the
UV and 24~\micron\ data and the average 70 and 160~\micron\ flux
densities from the stacking analyses,
SFR$_{\mathrm{UV}+24,70,160\micron}$.  The right and top axes show the
corresponding bolometric luminosity, $\lbol \equiv L_\mathrm{UV}
+ \lir$, under the assumption that star formation heats the dust
producing the IR emission.   Blue circles show ordinary IR sources
(without X-ray detections or IR--power-law-like colors).  Green
squares denote IR--power-law sources (excluding X-ray sources).  Red
pentagrams show X-ray sources.  The dashed line shows the unity
relation. The shaded region shows a polynomial fit to the data,
excluding the X-ray sources and IR--power-law sources (see text).  The
error bar shows typical errors on the derived SFRs, $\approx 0.5$~dex
for the abscissa and $0.1$~dex for the ordinate. \label{fig:sfrvsfr}}
\end{figure}

Under the assumption that star formation is primarily responsible for
the dust heating powering the IR emission, we estimate the
instantaneous SFR using the bolometric luminosity, defined as the sum
of the rest--frame UV and total IR luminosities of each galaxy,
$\Psi/1\; M_\odot = 1.8\times 10^{-10}
\times \lbol/\lsol$, where $\lbol \equiv (2.2 \times L_\mathrm{UV} +
\lir)$, \lir\ is the total IR luminosity and  $L_\mathrm{UV}$ is the
rest--frame UV luminosity derived from the monochromatic luminosity at
2800~\AA, $L_\mathrm{UV} = 1.5 \times L(2800\AAA)$, uncorrected for
extinction \citep[\eg,][]{bel05}.  The  SFR to bolometric luminosity
calibration assumes a Salpeter--like IMF with upper and lower mass
cutoffs of 0.1 and 100~\msol\ \citep[see,
\eg,][]{ken98,bel03,bel05}.   Figure~\ref{fig:sfrvsfr} compares the
SFRs derived using the 24~\micron\ data only to the SFRs derived from
the 24~\micron\ flux density with the average 70 and 160~\micron\ flux
densities.   Symbol type denotes objects with X-ray detections, and IR
power-law sources.

The relation in figure~\ref{fig:sfrvsfr} provides a tool for deriving
SFRs for $1.5 < z < 2.5$ galaxies ``corrected'' using the average 70
and 160~\micron\ flux densities given a bolometric luminosity derived
from the UV and 24~\micron\ flux.  Note, however, for the
24~\micron--selected sources $\lir$ constitutes the majority of the
bolometric luminosity, and $\lir(\mathrm{24,70,160\micron}) /
L_\mathrm{bol} = 0.85-1.0$ for all sources here.  We fit a
second--order polynomial to the data in figure~\ref{fig:sfrvsfr}
(excluding X-ray and IR--power-law sources) to derive an empirical
relation between the derived SFRs.  The fit is valid over the range
$\Psi \gsim 100$~\msol\ yr$^{-1}$ ($\log(L_\mathrm{bol}/L_\odot) \geq
11.8$).  The empirical relation is:
\begin{equation}
y = \sum_{i=0}^{2} C_i\, x^i, 
\end{equation}
where $y\equiv$ $\log ( \Psi_\mathrm{UV+24,70,160\micron})$, $x\equiv$
$\log (
\Psi_\mathrm{UV+24\micron}$), where $\Psi$ has units of \msol\
yr$^{-1}$, and  $\mathbf{C}$ = (0.37, 1.05, $-0.085$) are the
polynomial coefficients.   The standard error on $\log
(\mathrm{SFR}_\mathrm{UV+24,70,160\micron})$ is 0.1~dex, derived from
the polynomial fit.  The fit is indicated by the shaded band in the
figure.  A similar relation exists for the bolometric luminosities,
$\lbol$, replacing $\Psi$ in Equation~1 with, $x \equiv
\log\{\, [ 2.2\times L_\mathrm{UV}+\lir(24\micron)] / 5.6 \times 10^{9}\,\lsol
\}$ and $y\equiv  \log\{\, [ 2.2\times L_\mathrm{UV}+\lir(24,70,160\micron) /
5.6 \times 10^{9}\,\lsol \}$.  

Unsurprisingly, figure~\ref{fig:sfrvsfr}  illustrates that objects
with large bolometric luminosities --- and thus large SFRs ---
inferred from their 24~\micron\ flux densities are reduced
substantially when we include the average 70 and 160~\micron\ flux
densities from the stacking analysis.  As a result, some published
studies using the inferred SFRs and \lir\ for bright
24~\micron--selected objects at $z\sim 2$ deserve reexamination.
Several studies using data covering relatively small areas
($\sim$100--160~arcmin$^2$) found that sources at $z\sim 2$ have
average 24~\micron\ flux densities,
$\simeq$100--200~\ujy\ \citep{dad05,cap06,pap06,red06,web06}.  At
these flux densities our analysis shows that $\lir(24\micron)$ and
$\lir(24,70,160\micron)$ are comparable (within a factor 2 for the
DH02 models), and the average 70 and 160~\micron\ emission will have
little influence on the conclusions of these studies.   In contrast,
studies of 24~\micron\ sources at $z\sim 2$ over relatively larger
areas identify many more bright 24~\micron\ sources, where the average
70 and 160~\micron\ sources have the most affect.   For
example, \citet{per05} find that the characteristic luminosity of the
total IR luminosity function at $z=2$ is $\approx
4\times10^{12}$~\lsol\ (applying their scaling to the monochromatic
rest--frame 12~\micron\ luminosity), corresponding to $S_{24} \approx
300$~\ujy.  Our findings imply that IR luminosities of sources as
bright or brighter than the ``knee'' in this luminosity function will
be overestimated by factors of $>$2 (or greater than a factor of four
using the SK07 models).  Moreover, \citet{hou05} and \citet{yan07}
study the \spitzer/IRS spectra of $z\sim 2$ sources with $S_{24} >
0.75$~mJy and $>0.9$~mJy, respectively.  The total IR luminosities of
these sources will be overestimated by factors $>$3 (and possibly as
large as an order of magnitude) using solely on the 24~\micron\ flux
densities and local templates for star-forming galaxies.
Therefore, based on our stacking analysis at 70 and 160~\micron, the
bolometric luminosities and SFRs of bright 24~\micron\ galaxies at
$1.5 < z < 2.5$  are more modest than what is suggested based solely
on the 24~\micron\ flux densities.

Under the assumption that dust heating from star formation powers the
IR luminosities, the 24~\micron--derived SFRs for many sources are
$\gsim 1000$~\msol\ yr$^{-1}$, even after excluding X-ray and
IR--power-law sources.  Such systems are not observed locally, even in
IRAS--selected samples \citep[\eg,][]{san91,ken98}.  Including the
average 70 and 160~\micron\ flux densities from our stacking analysis
reduces the implied SFRs for bright 24~\micron\ sources at $1.5 < z <
2.5$ substantially, such that the vast majority of sources have  $\Psi
< 1000$~\msol~yr$^{-1}$ (see Figure~\ref{fig:sfrvsfr}).   Indeed,
sources with $\Psi > 1000$~\msol\ yr$^{-1}$ are very rare, with a
surface density of $30\pm10$~deg$^{-2}$, corresponding to $2\pm1\times
10^{-6}$~Mpc$^{-3}$ over $1.5 < z < 2.5$, and consistent with the
space density of sub-mm galaxies with comparable luminosities
\citep{cha05}.  Locally, IRAS--selected ULIRGs with SFRs of
many hundred solar masses per year have large gas masses,
$M(\mathrm{H}_2) \sim 10^{11}$~\msol, within small, centrally
concentrated radii, $r \lsim 2$~kpc (see Kennicutt 1998).   These
galaxies represent cases of a near--maximal rate of star formation
under physical and dynamical arguments, which convert all their gas
into stars in one dynamical time \citep[\eg,][]{len96}.   Current
measurements of the gas masses at high redshifts are limited to radio
galaxies, but show that large gas reservoirs abound
\citep[\eg,][]{gre04}.  Many distant star-forming galaxies
have small rest--frame optical sizes \citep{tru04,pap05}, comparable
to the IR--emitting regions in local ULIRGs, and such conditions may
exist in $z\sim 2$ galaxies with these high SFRs. However, there are
star-forming galaxies at $z\sim 2$ with extended sizes ($R \gsim
4$~kpc) \citep{lab03,zir07}, and clearly it is important to measure
the range of sizes of high--redshift IR--luminous galaxies directly.
Nevertheless, because the majority of high redshift galaxies in our
24~\micron--selected sample have $\lsim$1000~\msol\ yr$^{-1}$,  these
galaxies may have similar feedback and star formation efficiencies
comparable as lower redshift analogs.

Although local systems with star formation of this magnitude are
relatively rare, they are $\approx 1000$ times more common at
$z\approx 2$ \citep{dad05,pap06}.  As noted by \citet[see also Daddi
et al.\ 2007a]{dad05}, the high 24~\micron--detection fraction of
galaxies at $1.5 < z < 2.5$ implies that if the IR--active phases are
short lived (on the order of a dynamical time) they must undergo a
high duty cycle.  Thus, if a high redshift starburst consumes the
available gas mass, there must be an accompanying mechanism to
replenish the galaxies' gas supply and drive it to high densities
(e.g., cold--gas flows or successive galaxy mergers), in order to
maintain the high detection fraction.   Logically, the reservoir of
molecular gas in galaxies is substantially higher at $1.5 < z < 2.5$
than at present, and the enhanced fraction of systems with maximal
SFRs is likely a consequence of the fact that more gas is available to
fuel starbursts or AGN.  Thus, it may also be the case that the lower
number density of these systems at present is a consequence of the
fact that gas required to fuel IR--luminous stages of galaxy evolution
occurs with less frequency.

\section{Conclusions}

In this paper we explored the \spitzer\ 24, 70, and 160~\micron\
properties of high--redshift galaxies.  Our primary interest is to
improve the constraints on the total infrared (IR) luminosities of
these galaxies. We studied the 24 to 160~\micron\ flux--densities of
galaxies as a function of IR and X-ray activity  using
a \ks-band--selected sample of galaxies from the MUSYC data in the
ECDF--S. From $z \approx 0$ to 1.5, the majority of 70~\micron\ and
160~\micron--detected galaxies have flux--density ratios consistent
with local star-forming galaxies and AGN.  Only four galaxies at $1.5
< z < 2.5$ are detected at 70~\micron\ to the depth of the MIPS data
($> 4.6$~mJy, $3\sigma$ at 70~\micron), and these have high
$S_{70}/S_{24}$ flux-density ratios.

There are no galaxies with $1.5 < z < 2.5$, high 24~\micron\ flux
densities ($S_{24} > 0.6$~mJy)  detected directly at 70~\micron\ or
160~\micron.   Bright 24~\micron\ sources therefore have low
70~\micron\ flux densities,  $S_{70} < 4.6$~mJy.   This is not
expected if these bright 24~\micron\ sources have SEDs consistent with
expectations from empirical templates of local star-forming
IR--luminous galaxies.   While some bright 24~\micron\ sources at
these redshifts and flux densities with published mid--IR spectra have
strong PAH emission features, most have featureless spectra.
Therefore, the low $S_{70}/S_{24}$ ratios are not due solely to
boosted 24~\micron\ from PAHs.  It is more likely that their $S_{70}/S_{24}$ ratios result
from warm dust temperatures.

Because so few 24~\micron\ sources at $1.5 < z < 2.5$ are detected
directly in the longer wavelength \spitzer\ data,  we used a stacking
analysis to study the average 70 and 160~\micron\ flux density of
sources at this redshift as a function of 24~\micron\ flux density,
X-ray activity, and rest--frame near--IR color.

Ordinary IR sources at $1.5 < z < 2.5$ with $53 <
S_{24}/\ujy < 100$ and $100 < S_{24}/\ujy \leq 250$ have average flux
densities $S_{70}$=0.88 and 1.5~mJy, respectively.  The average flux
densities at 160~\micron\ are 4.6 and 5.1~mJy, respectively, although
these have low S/N ratios (1.8 and 2.3, respectively).   For these
galaxies the  $S_{70}/S_{24}$ ratio and the shape of the
observed--frame 24--160~\micron\ SED are generally consistent with
empirical models of IR--luminous galaxies.    This suggests that
\textit{on average} bolometric conversions from the measured 24~\micron\
flux density to total IR luminosities are fair.

Ordinary IR sources at $1.5 < z < 2.5$ with $S_{24} > 250$~\ujy\ have
average flux densities of $S_{70}$=2.0~mJy and $S_{160}$=4.6~mJy.
These sources have average $S_{70}/S_{24}$ ratios substantially lower
than predicted from our empirical models of local star-forming
galaxies, although they are similar to the local ULIRG IRAS 08572+3915
(Armus et al.\ 2007).  Theoretical model fits to the average 24, 70,
and 160~\micron\ flux densities have compact star-forming regions with
warm dust temperatures.   Such conditions may be expected
theoretically as high redshift galaxies with high gas densities and
small sizes similar to local ULIRGs should have high dust temperatures
\citep{kav03}.   Observationally, many star-forming galaxies at
high--redshifts have small optical sizes comparable to the
IR--emitting regions of local ULIRGs \citep{tru04,pap05} and we expect
the IR--emitting region to be no bigger than the optical radius.
Thus, if star formation powers high redshift sources with $S_{24} >
250$~\ujy, these objects generally are compact with warm dust
temperatures.  Deep high--angular--resolution imaging of these sources
either at rest--frame optical wavelengths (e.g., with \hst) or in the
far--IR (ALMA) are needed to measure the size distribution of these
sources.  Alternatively, ordinary IR sources with $S_{24} > 250$~\ujy\
at $1.5 < z < 2.5$ have $S_{70}/S_{24}$ ratios comparable to X-ray
sources with similar 24~\micron\ flux densities.  The $S_{70}/S_{24}$
ratios for these sources are also similar to the red tail of QSOs
distribution \citep{elv94}.  Therefore, there may be a significant AGN
contribution to the mid--IR colors of these galaxies.  Theoretical
model fits to the average 24, 70, and 160~\micron\ flux densities
shows they are consistent with dust heating from star formation or
AGN.  Moreover, they may also involve AGN/starburst composites.

We compared the total IR luminosities for $1.5 < z < 2.5$ galaxies
derived from the average 24, 70, and 160~\micron\ flux densities and
those inferred solely from the 24~\micron\ data.   For sources without
X-ray detections and $53 < S_{24}/\ujy < 100$ or $100 < S_{24}/\ujy <
250$,  the $\lir(24,70,160\micron)$  values are within factors of
$\approx$2--3 of $\lir(24\micron)$ (with a dependency on redshift and
on the choice of empirical or theoretical model).   However, sources
without X-ray detections and $S_{24} > 250$~\ujy\ have larger ratios
of $\lir(24\micron) / \lir(24,70,160\micron)$ ranging from factors of
$\approx$3--10 (again depending on redshift and the choice of model).
X-ray and IR--power-law sources have fairly substantial ratios,
ranging from factors of $2$ to $\approx 10$.  In all cases, the IR
luminosities derived solely from the 24~\micron\ data for bright
sources, $S_{24} > 250$~\ujy\ are overestimated, in some cases by large
factors.

We investigated how the average 70 and 160~\micron\ flux densities
affects the interpretation of the bolometric luminosities and SFRs in
galaxy samples at $1.5 < z < 2.5$ in the MUSYC data.   Using the
bolometric luminosities derived using the measured 24~\micron\ flux
densities, with the average 70, and 160~\micron\ flux densities, the
majority of 24~\micron--selected galaxies at $1.5 < \zph < 2.5$ have
IR luminosities $\lir \lsim 6 \times
10^{12}$~\lsol, which if attributed to star formation corresponds to
$\lsim 1000$~\msol\ yr$^{-1}$.    This is similar to the
maximal star formation rate observed in low redshift galaxies,
suggesting that high redshift galaxies may have similar star formation
efficiencies and feedback processes.   Objects with $\lir > 6\times
10^{12}$~\lsol\ are quite rare, with a surface density $\sim 30\pm
10$~deg$^{-2}$, corresponding to $\sim 2\pm 1\times10^{-6}$~Mpc$^{-3}$
over $1.5 < z < 2.5$.

\acknowledgments

We wish to thank our collaborators on the MUSYC and MIPS GTO teams for
many interesting discussions, and for their much hard work.  We also
thank the anonymous referee for their suggestions, which improved the
quality and clarity of this paper.  This work is also based in part on
data obtained with the Spitzer Space Telescope, which is operated by
the Jet Propulsion Laboratory (JPL), California Institute of
Technology (Caltech) under a contract with NASA. Support for this work
was provided by NASA through the Spitzer Space Telescope Fellowship
Program, through a contract issued by JPL, Caltech  under a contract
with NASA.

\appendix

\section{Stacking \textit{Spitzer} MIPS Data}\label{section:appendix}

\begin{figure}[tp]
\vspace{-24pt}
\epsscale{1.1}
\plottwo{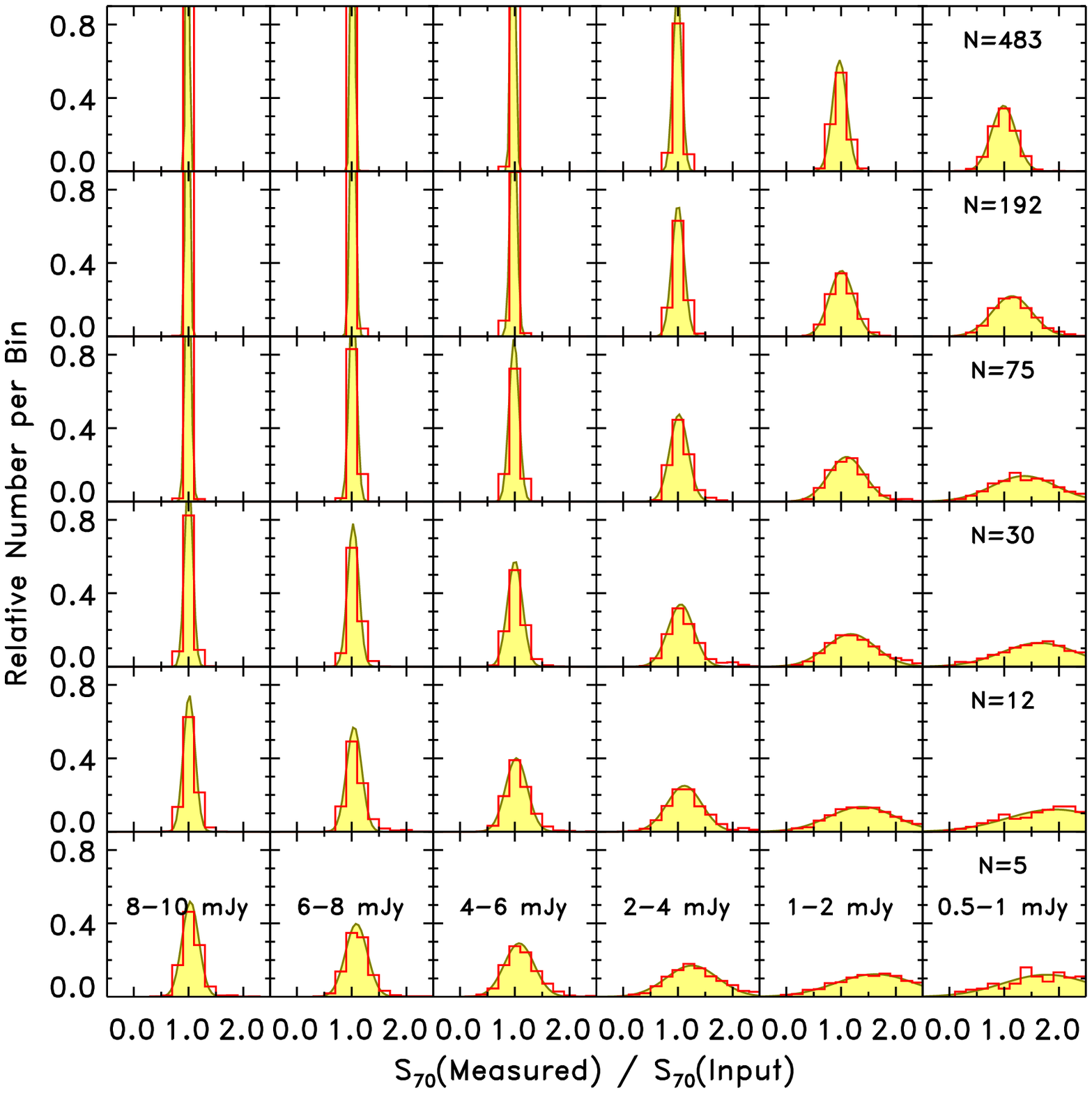}{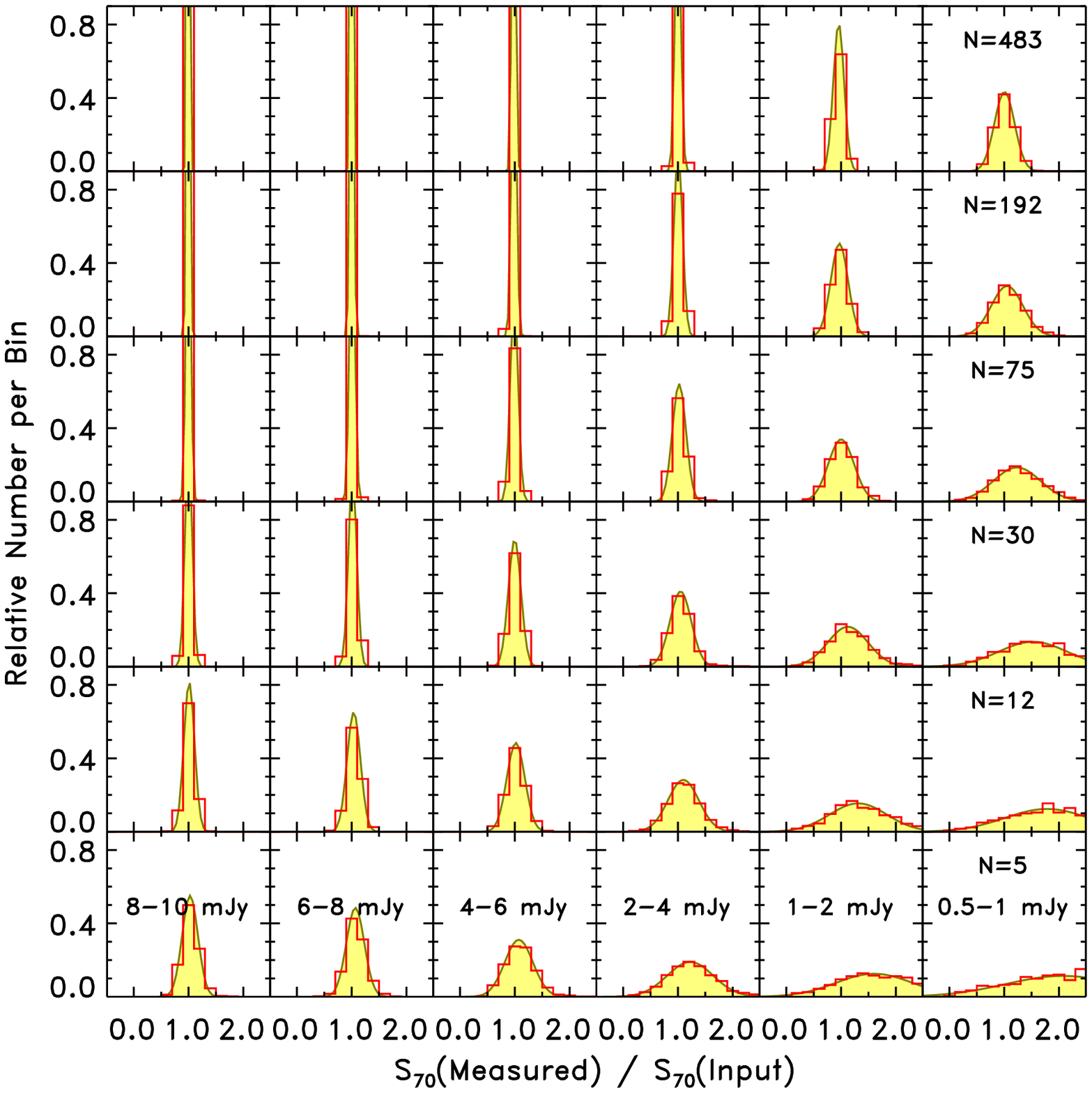}
\vspace{-0.2in}
\epsscale{0.667}
\plotone{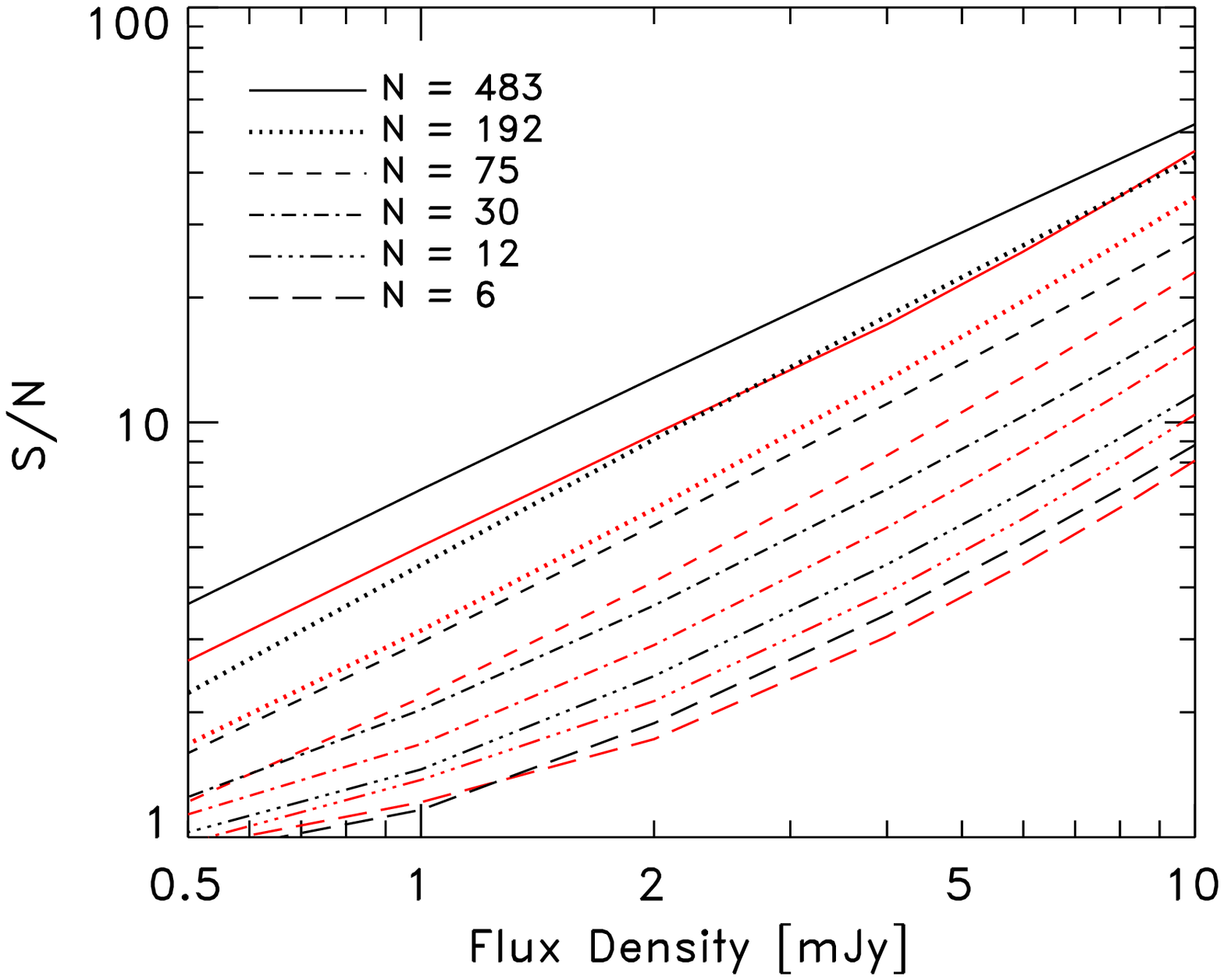}
\epsscale{1.0}
\caption{  The top panels show the distribution of
recovered 70~\micron\ flux densities from stacking simulations.  Each
sub--panel shows the distribution of the ratio of measured 70~\micron\
flux density to input flux density.  Each column shows this
distribution as a function of measured 70~\micron\ flux density (as
labeled), and each row shows this distribution as a function of the
number of sources in the stack, $N$ (as labeled).   The left top panel
shows the results from stacking sources in the original image
containing all 70~\micron\ sources.  The right top panel shows the
results from stacking sources in images that have had sources detected
at 70~\micron\ with $>$5$\sigma$ significance removed.   The bottom
panel shows the signal--to--noise for the simulated stacked images as
a function of flux density, where we derive the signal--to--noise
using the width of the distributions as an estimate of the flux
density error.   The line--types show the S/N as a function of $N$
objects in each stack, as labeled in the figure inset.  Black lines
correspond to stacked images that have had bright sources removed.
Red lines correspond to stacked images without removing
sources. \label{fig:appendix}}
\end{figure}

The vast majority of \spitzer/MIPS 24~\micron\ sources with $1.5 < z <
2.5$ are undetected to the flux density limits of the  MIPS 70 and 160~\micron\
data.   To study the far--IR properties of these sources, we
resort to stacking the 70 and 160~\micron\ data at the positions of the
24~\micron\ sources.  Several studies have illustrated the benefit of
stacking sources in MIPS data \citep{zhe06,zhe07,dol06,huy07,dye07}.  In this
section we describe our stacking methodology, adapted from these prior
works.

The first step in the stacking procedure is to take a small subimage
from the 70 and 160~\micron\ data centered at the astrometric position
of each 24~\micron\ source to be stacked.  For this step, we use
subimages of approximately $200\arcsec \times 200\arcsec$ for both the
70 and 160~\micron\ data.   The subimage size does not affect the
average (stacked) value so long as it encompasses sufficient area for a local
background measurement to be made on the stacked source.    We use a
two--dimensional bilinear interpolation to center the 70 and
160~\micron\ subimages on the astrometric coordinates of the
24~\micron\ source.   We then subtract the local background measured
in annuli of 39--65\arcsec\ and 32--56\arcsec\ of each 70 and
160~\micron\ subimage, respectively.

In the second step we sum the images and take the mean to derive the
average flux density of the sample.   Following \citet{dol06} and
\citet{huy07}, we rotate each subimage by 90$^{\circ}$ relative to the
previous subimage to reduce the effects of image artifacts on the
average measurement.  We experimented using other combination schemes,
including taking the median of each pixel in the stack or taking the
mean after rejecting outliers.  In general, we found these to all be
consistent once we removed sources detected in the 70 and 160~\micron\
(see below).  Thus, the method used to combine the subimages is
secondary relative to ensuring that the subimages are clean of bright sources.

We tested our stacking method using simulations of artificial sources
of known flux density randomly placed in the ECDF--S 70 and
160~\micron\ images.   We compared the accuracy of the flux density
measured in the stacked image to the true value.   We performed 10,000
Monte Carlo simulations for each bin of flux density and number of
objects to measure reliable statistics as a function of flux density
and number of objects in the stack.  Figure~\ref{fig:appendix} shows
the distribution of the ratio of the measured flux density to the true
(input) flux density as a function of flux density and number of
sources  for the 70~\micron\ simulations.  The width of the
distribution increases with decreasing 70~\micron\ flux density, and
decreasing number of objects each stack.  We find similar
behavior in the distribution for the 160~\micron\ simulations.

The mean of the measured--to--input flux density ratio shifts to
values greater than one for low numbers of objects and fainter flux
density.   This is a consequence of two effects.  One is that the flux
from nearby detected objects bias the measured flux density.  The
other is a result of ``confusion noise'' from nearby unresolved
sources, which contribute to the average stacked value (and similar to
effect in confusion--limited sub--mm photometry discussed by Coppin et
al.\ 2005).    To suppress the effect of detected sources from
contributing to the stacking, we repeated our Monte Carlo test after
first removing sources with $>$5$\sigma$
detections in the 70 and 160~\micron\ images.   The top right panel in
figure~\ref{fig:appendix} shows the distribution of
measured--to--input flux values using the source--subtracted images
for the 70~\micron\ simulations before stacking.

We found appreciable gains in the accuracy of the average value for
images cleaned of sources detected in the 70 and 160~\micron\ images.
We derive the error, $\sigma$, on the stacking measurement using the
width of the distribution of measured--to--input flux ratio.   The
bottom panel of figure~\ref{fig:appendix} illustrates the gain in
signal--to--noise ($= f_\nu / \sigma$) measured from
the stacked images as a function of flux density and number of objects
in each stack.  For example, for an average measurement of 100
simulated objects with $S_{70} = 1$~mJy the uncertainty is
0.42~mJy without excluding sources. The uncertainty drops to 0.30~mJy
with sources excluded.   We therefore use 70 and 160~\micron\ cleaned
of sources for our stacking analysis.

\end{document}